\documentclass[aps,pra,twocolumn,superscriptaddress, nofootinbib]{revtex4-1}
\pdfoutput=1
\usepackage[utf8]{inputenc}
\usepackage[english]{babel}
\usepackage[T1]{fontenc}
\usepackage{amsmath}
\usepackage{hyperref}
\usepackage{tikz}
\usepackage{lipsum}
\usepackage{float}
\usepackage{graphicx}
\usepackage{latexsym}
\usepackage{amssymb}
\usepackage{nicefrac}
\usepackage{amsfonts}
\usepackage{enumerate}
\usepackage{upgreek}
\usepackage{subfigure}
\usepackage{bm}
\usepackage[version=3]{mhchem}
\usepackage{nicefrac}
\usepackage{bbold}
\usepackage{verbatim}
\usepackage{multirow}
\usepackage{color}
\usepackage{comment}
\usepackage{xcolor}
\usepackage{soul}

\usepackage{braket}

\usepackage{tikz}
\newcommand{\singlet}{
  \begin{tikzpicture}[baseline=-0.5ex]
  \begin{scope}[scale=0.8]
    \draw (0,0) ellipse (0.1 and 0.25);
  \end{scope}
  \end{tikzpicture}
}

\newcommand{\holesigma}{
  \begin{tikzpicture}[baseline=-0.5ex]
  \begin{scope}[scale=0.8]
    \node[label={{$\sigma$}}] at (0,-0.5) {};
    \draw (0,0.2) circle (0.1);
  \end{scope}
  \end{tikzpicture}
}

\newcommand{\downup}{
  \begin{tikzpicture}[baseline=-0.5ex]
  \begin{scope}[scale=0.8]
    \node[label={{$\uparrow$}}] at (0,-0.7) {};
    \node[label={{$\downarrow$}}] at (0,-0.2) {};
  \end{scope}
  \end{tikzpicture}
}

\newcommand{\upup}{
  \begin{tikzpicture}[baseline=-0.5ex]
  \begin{scope}[scale=0.8]
    \node[label={{$\uparrow$}}] at (0,-0.7) {};
    \node[label={{$\uparrow$}}] at (0,-0.2) {};
  \end{scope}
  \end{tikzpicture}
}

\newcommand{\holehole}{
  \begin{tikzpicture}[baseline=-0.5ex]
  \begin{scope}[scale=0.8]
    \draw (0,0.2) circle (0.1);
    \draw (0,-0.1) circle (0.1);
  \end{scope}
  \end{tikzpicture}
}
\newcommand{\holeup}{
  \begin{tikzpicture}[baseline=-0.5ex]
  \begin{scope}[scale=0.8]
   \draw (0,0.2) circle (0.1);
    \node[label={{$\uparrow$}}] at (0,-0.7) {};
  \end{scope}
  \end{tikzpicture}
}
\newcommand{\holedown}{
  \begin{tikzpicture}[baseline=-0.5ex]
  \begin{scope}[scale=0.8]
   \draw (0,0.2) circle (0.1);
    \node[label={{$\downarrow$}}] at (0,-0.7) {};
  \end{scope}
  \end{tikzpicture}
}

\newcommand{\C}{\hat{c}}
\newcommand{\Cd}{\hat{c}^\dagger}

\newcommand{\Bd}{\hat{b}^\dagger}
\newcommand{\B}{\hat{b}}

\newcommand{\n}{\hat{n}}
\newcommand{\Ham}{\hat{\mathcal{H}}}
\newcommand{\hc}{\mathrm{h.c.}}

\renewcommand{\vec}[1]{\mathbf{#1}}

\begin{document}

\title{Local control and mixed dimensions: Exploring high-\\temperature  superconductivity in optical lattices}

\author{Henning Schl\"omer}
\affiliation{Department of Physics and Arnold Sommerfeld Center for Theoretical Physics (ASC), Ludwig-Maximilians-Universit\"at M\"unchen, Theresienstr. 37, M\"unchen D-80333, Germany}
\affiliation{Munich Center for Quantum Science and Technology (MCQST), Schellingstr. 4, D-80799 M\"unchen, Germany}
\affiliation{Department of Physics, Harvard University, Cambridge MA 02138, USA}
\author{Hannah Lange}
\affiliation{Department of Physics and Arnold Sommerfeld Center for Theoretical Physics (ASC), Ludwig-Maximilians-Universit\"at M\"unchen, Theresienstr. 37, M\"unchen D-80333, Germany}
\affiliation{Munich Center for Quantum Science and Technology (MCQST), Schellingstr. 4, D-80799 M\"unchen, Germany}
\affiliation{Max-Planck-Institute for Quantum Optics, Hans-Kopfermann-Str.1, Garching D-85748, Germany}
\author{Titus Franz}
\affiliation{Munich Center for Quantum Science and Technology (MCQST), Schellingstr. 4, D-80799 M\"unchen, Germany}
\affiliation{Max-Planck-Institute for Quantum Optics, Hans-Kopfermann-Str.1, Garching D-85748, Germany}
\author{Thomas Chalopin}
\affiliation{Laboratoire Charles Fabry, Institut d'Optique Graduate School, CNRS, Universit\'e Paris-Saclay, 91127 Palaiseau, France}
\author{Petar Bojovi\'c}
\affiliation{Munich Center for Quantum Science and Technology (MCQST), Schellingstr. 4, D-80799 M\"unchen, Germany}
\affiliation{Max-Planck-Institute for Quantum Optics, Hans-Kopfermann-Str.1, Garching D-85748, Germany}
\author{Si Wang}
\affiliation{Munich Center for Quantum Science and Technology (MCQST), Schellingstr. 4, D-80799 M\"unchen, Germany}
\affiliation{Max-Planck-Institute for Quantum Optics, Hans-Kopfermann-Str.1, Garching D-85748, Germany}
\author{Immanuel Bloch}
\affiliation{Department of Physics and Arnold Sommerfeld Center for Theoretical Physics (ASC), Ludwig-Maximilians-Universit\"at M\"unchen, Theresienstr. 37, M\"unchen D-80333, Germany}
\affiliation{Munich Center for Quantum Science and Technology (MCQST), Schellingstr. 4, D-80799 M\"unchen, Germany}
\affiliation{Max-Planck-Institute for Quantum Optics, Hans-Kopfermann-Str.1, Garching D-85748, Germany}
\author{Timon A. Hilker}
\affiliation{Munich Center for Quantum Science and Technology (MCQST), Schellingstr. 4, D-80799 M\"unchen, Germany}
\affiliation{Max-Planck-Institute for Quantum Optics, Hans-Kopfermann-Str.1, Garching D-85748, Germany}
\author{Fabian Grusdt}
\affiliation{Department of Physics and Arnold Sommerfeld Center for Theoretical Physics (ASC), Ludwig-Maximilians-Universit\"at M\"unchen, Theresienstr. 37, M\"unchen D-80333, Germany}
\affiliation{Munich Center for Quantum Science and Technology (MCQST), Schellingstr. 4, D-80799 M\"unchen, Germany}
\author{Annabelle Bohrdt}
\affiliation{Munich Center for Quantum Science and Technology (MCQST), Schellingstr. 4, D-80799 M\"unchen, Germany}
\affiliation{Institut für Theoretische Physik, Universität Regensburg, D-93035 Regensburg, Germany}

\date{\today}
\begin{abstract}
The simulation of high-temperature superconducting materials by implementing strongly correlated fermionic models in optical lattices is one of the major objectives in the field of analog quantum simulation. Here we show that local control and optical bilayer capabilities combined with spatially resolved measurements create a versatile toolbox to study fundamental properties of both nickelate and cuprate high-temperature superconductors. On the one hand, we present a scheme to implement a mixed-dimensional (mixD) bilayer model that has been proposed to capture the essential pairing physics of pressurized bilayer nickelates. This allows for the long-sought realization of a state with long-range superconducting order in current lattice quantum simulation machines. In particular, we show how coherent pairing correlations can be accessed in a partially particle-hole transformed and rotated basis. On the other hand, we demonstrate that control of local gates enables the observation of $d$-wave pairing order in the two-dimensional (single-layer) repulsive Fermi-Hubbard model through the simulation of a system with attractive interactions. Lastly, we introduce a scheme to measure momentum-resolved dopant densities, providing access to observables complementary to solid-state experiments -- which is of particular interest for future studies of the enigmatic pseudogap phase appearing in cuprates. 
\end{abstract}
\maketitle

\section{Introduction}
Though the discovery of high-temperature superconductivity in copper oxide compounds dates back almost 40 years~\cite{Bednorz1986}, a full understanding of its phase diagram at finite doping remains elusive. The paradigmatic two-dimensional (2D) Fermi-Hubbard (FH) model, believed to capture the essential low-energy physics of high-$T_c$ compounds~\cite{Anderson1987, Lee2006, Keimer2015}, has been subject to particular theoretical and experimental scrutiny in the past decades. Advances in numerical techniques have allowed to shed new light on the intricate competition between various strongly correlated phases in the ground state of the FH model~\cite{LeBlanc2015, Zheng2017, Jiang2019_science, Qin_absence_SC, MultiMess2021, Jiang_Kivelson, Arovas2022_rev, xu2023coexistence}. Furthermore, technological innovations in the field of analog quantum simulation~\cite{Bloch2008, Esslinger2010, Bloch2012, Parsons2015, Cheuk2015, Haller2015, Gross2017, Bohrdt2020} have led to a plethora of insights into the intermediate-temperature regime by analyzing real-space correlations, including the observation of magnetic order~\cite{Greif2013, Hart2015, Boll2016, Mazurenko_AFM} and the formation of magnetic polarons~\cite{Koepsell_nature2019, Chiu2019, Hartke2020, Koepsell2021, Ji2021}. With quantum gas microscope techniques, spin-resolved density measurements of the many-body state in the Fock basis can be obtained. However, the following significant challenges arise when attempting to microscopically study high-temperature superconducting (and their exotic normal) phases: $(a)$ Superconducting order of quantum many-body states can not be accessed using local densities, as corresponding observables are off-diagonal in the Fock basis. In particular, this requires adding and removing singlet pairs to the system in a coherent fashion, which is an exceptional challenge with currently available techniques. Furthermore, small energy differences between various collectively ordered phases lead to small critical temperatures~\cite{Zheng2017, Wietek_stripes, Qin_absence_SC}, which are out of reach for state-of-the-art quantum simulators. $(b)$ For the study of the exotic normal phases in cuprate superconductors, momentum-resolved observables of dopants are highly relevant~\cite{Lee2006, Norman_pseudogap, Chowdhury_pseudogap}. Though momentum-space densities of particles can be accessed using time of flight measurements, the latter is not possible for the dopants of the system.

Recently, mixed-dimensional (mixD) systems~\cite{Grusdt_tJ, Grusdt_tJz, bohrdt2021strong, Schloemer2022, Schloemer2022_recon} that can be engineered in optical lattices subject to potential gradients~\cite{Duan2003, Trotzky, Dimitrova} have emerged as a compelling tool to energetically favor and study collective phenomena in strongly correlated models. In particular, this allowed for the observation of real-space hole pairing due to magnetic correlations in tailored ladder geometries~\cite{Hirthe2022}, and provided novel insights into the formation of stripes on the 2D square lattice~\cite{bourgund2023}. The experimental setup to engineer mixD systems consists of an optical superlattice, i.e. coupled double wells, with tunable energy offset $\Delta$ [see e.g. the left-hand side of Fig.~\ref{fig:V} for an illustration]. This allows for the realization of meta-stable states where the hopping along the potential gradient is suppressed, while spin-exchange remains finite~\cite{Duan2003, Trotzky, Dimitrova}. 

With the discovery of high-temperature superconductivity at $T_c \sim 80$ K in pressurized bilayer nickelates~\cite{Sun2023, zhang2023_zeroR, hou2023emergence}, mixD systems have gained broad attention also in the condensed matter community. Specifically, mixed dimensions are widely believed to play an essential role in the formation of superconductivity in the bilayer nickelate La$_3$Ni$_2$O$_7$ (LNO)~\cite{Luo2023, lu2023, oh2023type, qu2023}. Indeed, simulations of minimal, single-band models suggest astonishingly high critical temperatures of the order of the magnetic coupling $J_{\perp}/2$ in certain parameter regimes~\cite{schlömer2023superconductivity}, which are readily achievable in state-of-the-art quantum simulation experiments~\cite{Hirthe2022}. 

\begin{figure*}
\centering
\includegraphics[width=\textwidth]{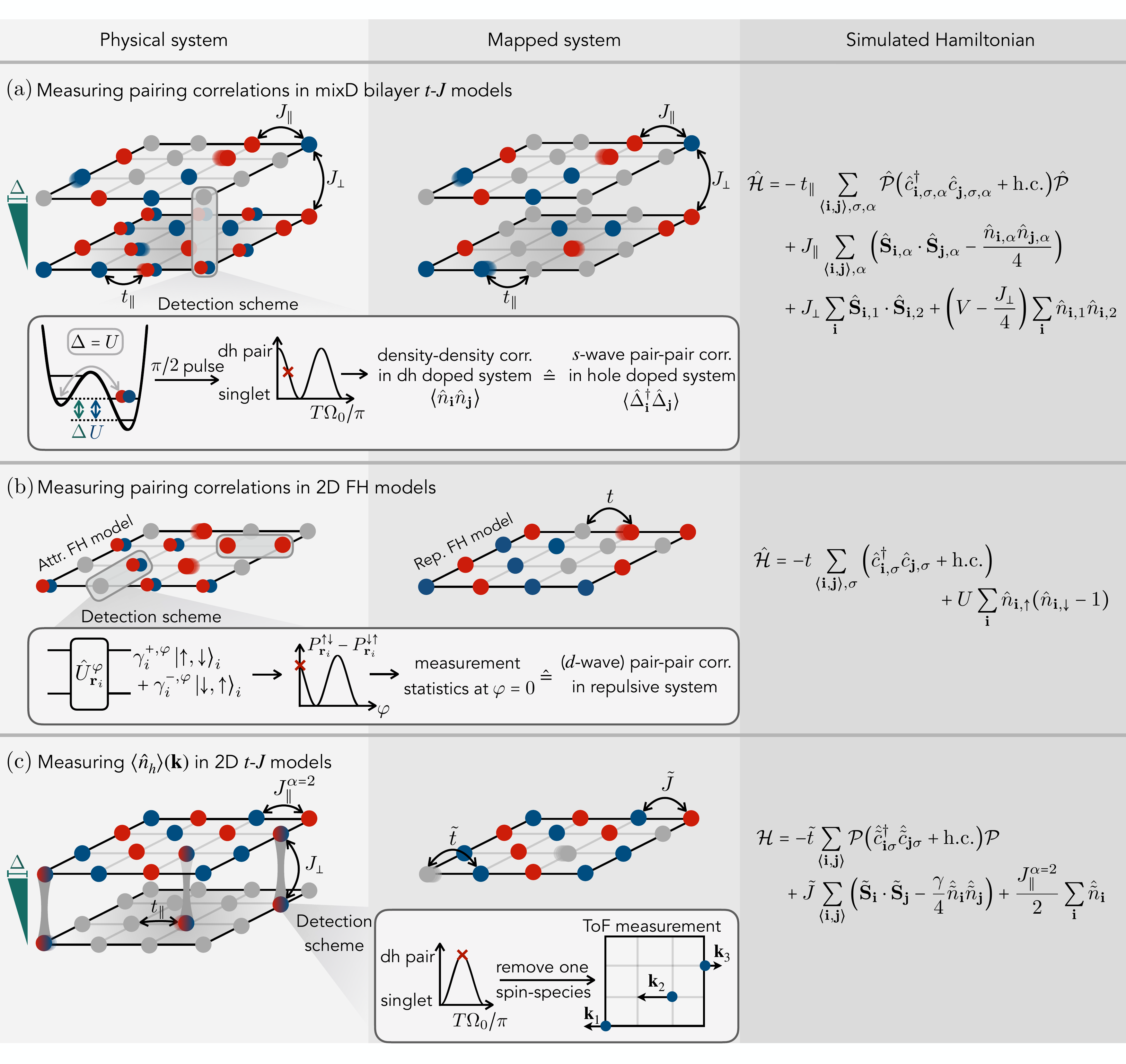}
\caption{\textbf{Exploring high-temperature superconductivity with optical lattices.} (a) Scheme how to measure coherent pairing correlations in the mixD bilayer $t-J$ model. We propose to implement the Fermi-Hubbard model in the strong-coupling limit on a bilayer geometry (physical system). By applying a potential gradient $\Delta$ between the two layers, interlayer tunneling is suppressed, such that $t_{\perp} = 0$, but $J_{\perp} > 0$. Up to tunable interlayer density-density interactions, the model relates to a fully hole-doped mixD bilayer $t$-$J$ model through a partial particle-hole transformation (mapped system). In the physical, doublon-hole-doped system, performing a global resonant tunneling pulse ($\Delta = U$) after ramping up the lattice depth results in Rabi oscillations with frequency $\Omega_0$ between singlets and doublon-hole (dh) pairs on the rungs. In a $\pi/2$ rotated basis (red cross), density-density correlations in the doublon-hole-doped system then correspond to coherent pair-pair correlations in the fully hole-doped system. The protocol is described in Sec.~\ref{sec:mixDbilayer}. (b) In a similar spirit, the 2D repulsive FH model relates to its attractive counterpart by a partial particle-hole transformation. To measure pairing correlations of various symmetries (including $d$-wave) $\langle \hat{\Delta}^\dagger_{\mathbf{r}_1}\hat{\Delta}_{\mathbf{r}_2}\rangle$ between bonds $\mathbf{r}_1$ and $\mathbf{r}_2$, we propose to implement the attractive FH model and apply local gates with unitaries $\hat{U}_{\mathbf{r}_i}^{\varphi}$ acting on bonds $\mathbf{r}_i$. This gives rise to a Ramsey interferometer as a function of a controlled phase dependency $\varphi$, from which pairing correlations can be accessed in the repulsive model. In particular, probabilities $P_{\mathbf{r}_i}^{\sigma_1 \sigma_2}$ of measuring $\ket{\uparrow, \downarrow}$ and $\ket{\downarrow, \uparrow}$ states after applying the local gates correspond to pairing correlations in the repulsive model. The protocol is described in Sec.~\ref{sec:FH_pairing} (c) By preparing the upper layer of a mixD setup at half-filling and loading (particle) dopants into the (initially empty) lower layer, singlets can be mapped to holes in an effective 2D $t$-$J$ model for $J_{\perp}\gg t_{\parallel}, J_{\parallel}^{\alpha=1,2}$. Singlet-to-dh transitions then give access to momentum-resolved hole densities via time-of-flight measurements in the lower layer. The protocol is described in Sec.~\ref{sec:tJ}.}
\label{fig:overview}
\end{figure*}

In this article, we present how local control of gates and bilayer optical lattice capabilities can be independently utilized to simulate minimal models and measure observables relevant to both nickelate and cuprate high-temperature superconductors. In particular, we argue that these methods overcome the challenges in microscopically studying superconducting and exotic normal phases: We show how current state-of-the-art quantum simulators can be used to $(i)$ prepare and observe a state with superconducting order, i.e. (quasi) long-range pair coherence, at realistic temperatures in the mixD bilayer $t$-$J$ model, $(ii)$ measure $d$-wave pairing correlations in the 2D FH model, and $(iii)$ access momentum-resolved dopant distribution functions in the 2D $t$-$J$ model. This directly facilitates complementary measurements to solid-state experiments of both bilayer nickelate and cuprate high-temperature superconductors using analog quantum simulation. The simulated Hamiltonians for proposals $(i)-(iii)$ are shown on the right-hand side of Fig.~\ref{fig:overview}. 

In the context of nickelate superconductors, we present a scheme to simulate the 2D mixD bilayer $t$-$J$ model on the square lattice and adiabatically prepare states that feature quasi long-range pairing correlations, 
\begin{equation}
\braket{\hat{\Delta}^{\dagger}_{\mathbf{i}} \hat{\Delta}_{\mathbf{j}}} \simeq \begin{cases}
        |\mathbf{i} - \mathbf{j}|^{-\alpha} \quad 0<T<T_c \\
        \text{const} \quad \quad \, \, T = 0,
    \end{cases}
\end{equation}
where $\hat{\Delta}_{\mathbf{i}}^{\dagger} = \frac{1}{\sqrt{2}} \left(\hat{c}^{\dagger}_{\mathbf{i}, \uparrow, \alpha = 1} \hat{c}^{\dagger}_{\mathbf{i}, \downarrow, \alpha = 2} - \hat{c}^{\dagger}_{\mathbf{i}, \downarrow, \alpha = 1} \hat{c}^{\dagger}_{\mathbf{i}, \uparrow, \alpha = 2}\right)$ creates an interlayer singlet between layers $\alpha = 1,2$ on site $\mathbf{i}$. The experimental setup consists of two FH layers with energy offset $\Delta$ in the strong-coupling limit, giving rise to interlayer magnetic interactions, as well as intralayer tunneling and magnetic coupling; hopping of particles between the two layers, however, is suppressed, making the system mixed-dimensional~\cite{Trotzky, Dimitrova, Hirthe2022, bourgund2023}. The setup is summarized in  Fig.~\ref{fig:overview}~(a). An essential ingredient that allows for the measurement of pair-pair correlations is to hole-dope one layer, while doublon-doping the other layer~\cite{Lange2023_1, Lange2023_2}, see the left-hand side of Fig.~\ref{fig:overview}~(a). In Sec.~\ref{sec:mixDbilayer}, we start by showing that, on bipartite lattices, the doublon-hole-doped bilayer mixD $t$-$J$ system is equivalent (up to tunable interlayer density-density interactions) to a fully hole-doped description by a partial particle-hole transformation applied to one layer only [see the right-hand side of Fig.~\ref{fig:overview}~(a)]. 

Using the density matrix renormalization group (DMRG), we calculate Luttinger exponents of pair-pair correlations in the ground state of the effective mixD model on a ladder and demonstrate that quasi long-range pairing order exists for a broad range of experimentally relevant parameters. Subsequently, we present a minimal adiabatic preparation scheme of a quantum state featuring pair-coherence. We further propose a measurement protocol involving resonant global interlayer tunneling $\pi/2$ pulses, which allows to access superconducting (pair-pair) correlations in the particle-hole transformed Hamiltonian. These correlations map to density-density correlations in the physically implemented doublon-hole-doped system and are hence readily accessible, without requiring to change the number of fermions. We propose to apply this scheme also to experimentally accessible 2D mixD bilayers, in which the Berezinskii–Kosterlitz–Thouless (BKT) transition to a superconducting state with quasi long-range pairing correlations around $T_c \sim J_{\perp}/2$ can be explored~\cite{schlömer2023superconductivity}. 

In connection with cuprates, in Sec.~\ref{sec:FH_pairing} we present a related scheme that allows to measure coherent pairing correlations in the 2D FH model on the square lattice. Following the ideas of Ho, Cazalilla and Giamarchi~\cite{Ho2009}, we consider an implementation of the FH model with strong attractive interactions~\cite{Hartke2023}, which is equivalent to the repulsive system through a partial particle-hole transformation, see Fig.~\ref{fig:overview}~(b). Coherent pairing order in the repulsive FH model can then be accessed through local basis-rotations in the implemented (attractive) model. In particular, we show that local control of tunneling gates allows for the observation of pairing correlations with different symmetries, e.g., the state can be probed on both $s$-wave and $d$-wave pairing order. Thereby we extend the ideas from Ref.~\cite{Ho2009}, where noise-correlation measurements have been proposed to analyze the antiferromagnetic state on the attractive side. With recent advances in local control in optical lattices~\cite{impertro2023local}, our scheme paves the way for the long-sought demonstration of $d$-wave pairing correlations in the plain-vanilla Hubbard model.

The toolbox of doped mixD bilayers additionally allows for the exploration of momentum-resolved observables of mobile holes in 2D $t$-$J$ models, summarized in Fig.~\ref{fig:overview}~(c). In particular, in Sec.~\ref{sec:tJ} we present a protocol to measure the free-hole (dopant) density in an effective 2D $t$-$J$ model in momentum-space, $\braket{\hat{n}_h(\mathbf{k})}$, which is a particularly relevant observable for revealing the properties of exotic normal phases (such as the appearance of a small Fermi surface in the pseudogap phase) of cuprates~\cite{Lee2006, Norman_pseudogap, Chowdhury_pseudogap}. We note that this is in contrast to direct implementations of the hole-doped 2D $t$-$J$ model, which give access to momentum resolved particle - but not dopant densities. 

To this end, we propose to implement a mixD bilayer system in the limit of strong interlayer Kondo-type couplings, i.e. strong spin exchange $J_\perp$ without tunneling $t_\perp$ between the layers, where mobile singlets can be mapped to holes in an effective 2D model, Fig.~\ref{fig:overview}~(c). By coherently driving tunneling transitions between interlayer singlets and doublon-hole pairs and the subsequent removal of one spin species, the momentum distribution of dopants $\braket{\hat{n}_h(\mathbf{k})}$ can be accessed through time-of-flight measurements. In particular, $\braket{\hat{n}_h(\mathbf{k})}$ does not depend on the spectral weight at the respective momentum $\mathbf{k}$, which can be advantageous compared to angle-resolved photoemission spectroscopy (ARPES) in regions of the Brillouin zone with low spectral weight, such as the backside of the Fermi arcs \cite{Shen2005}. Furthermore, the effective 2D $t$-$J$ model features hopping and spin interaction amplitudes that originate from different layers. This allows for an independent tuning of these parameters, and hence to simulate regimes that cannot be accessed through a direct implementation of a 2D layer.

\section{Measuring pairing correlations: Mixed-dimensional bilayers}
\label{sec:mixDbilayer}
In the following, we present how states with quasi long-range superconducting order can be prepared and how coherent pairing correlations can be measured in realistic experimental setups by implementing the mixD bilayer $t$-$J$ model in a transformed basis. 

An essential ingredient to measure pairing correlations in the mixD bilayer $t$-$J$ model is to experimentally implement a partially particle-hole transformed Hamiltonian. Therefore, before precisely defining the proposed model, we review the particle-hole symmetry of the standard $t$-$J$ model on bipartite lattices, retrieved from perturbation theory from the FH model. We note that relevant models for bilayer nickelates are defined on the (bipartite) square lattice, where the particle-hole symmetry and hence our proposed measurement scheme holds.

\subsection{Particle-hole symmetry of the conventional $t$-$J$ model}
When hole-doping the FH model away from one particle per site and projecting out states with double occupancy (valid in the strongly interacting limit $U \gg t$), the Hamiltonian reads (neglecting three-site, next-nearest neighbor terms $\sim J$)
\begin{equation}
\begin{aligned}
    \hat{\mathcal{H}} = -t \sum_{ \braket{\mathbf{i}, \mathbf{j}}, \sigma} \hat{\mathcal{P}} \big(&\hat{c}_{\mathbf{i}, \sigma}^{\dagger} \hat{c}_{\mathbf{j}, \sigma}^{\vphantom\dagger} + \text{h.c.} \big)\hat{\mathcal{P}} \\ &+ J \sum_{\braket{\mathbf{i}, \mathbf{j}}} \Big( \hat{\mathbf{S}}_{\mathbf{i}} \cdot \hat{\mathbf{S}}_{\mathbf{j}} - \frac{\hat{n}_{\mathbf{i}}\hat{n}_{\mathbf{j}}}{4} \Big).
\end{aligned}
\label{eq:tJh}
\end{equation}
Here, $\hat{c}_{\mathbf{i}, \sigma}^{(\dagger)}$, $\hat{n}_{\mathbf{i}}$ and $\hat{\mathbf{S}}_{\mathbf{i}}$ are fermionic annihilation (creation), particle density, and spin operators on site $\mathbf{i}$, respectively; $\braket{\mathbf{i}, \mathbf{j}}$ denotes nearest neighbor (NN) sites on the two-dimensional (2D) square lattice, and $\hat{\mathcal{P}}$ is the Gutzwiller operator projecting out states with double occupancy,
\begin{equation}
    \hat{\mathcal{P}} = \prod_{\mathbf{i}} \left( 1 - \hat{n}_{\mathbf{i}, \uparrow} \hat{n}_{\mathbf{i}, \downarrow} \right).
\end{equation}
The total particle number is given by $\hat{N} = \sum_{\mathbf{i}} \hat{n}_{\mathbf{i}, \uparrow} + \hat{n}_{\mathbf{i}, \downarrow} = L - d$, where $L$ and $d$ are the number of sites and (hole) dopants in the system, respectively. 

Similarly, we can consider doublon-doping the FH model. The perturbation theory works identically, however now we project out empty states, denoted by the projector $\hat{\widetilde{\mathcal{P}}}$,
\begin{equation}
    \hat{\widetilde{\mathcal{P}}} = \prod_{\mathbf{i}} \Big( 1 - (1-\hat{n}_{\mathbf{i}, \uparrow})(1- \hat{n}_{\mathbf{i}, \downarrow}) \Big) .
\end{equation}
Up to an overall doping dependent energy shift due to double occupancies, the Hamiltonian reads
\begin{equation}
\begin{aligned}
    \hat{\widetilde{\mathcal{H}}} = -t \sum_{ \braket{\mathbf{i}, \mathbf{j}}, \sigma} \hat{\widetilde{\mathcal{P}}} \big(&\hat{c}_{\mathbf{i}, \sigma}^{\dagger} \hat{c}_{\mathbf{j}, \sigma}^{\vphantom\dagger} + \text{h.c.} \big)\hat{\widetilde{\mathcal{P}}} \\ &+ J \sum_{\braket{\mathbf{i}, \mathbf{j}}} \Big( \hat{\mathbf{S}}_{\mathbf{i}} \cdot \hat{\mathbf{S}}_{\mathbf{j}} - \frac{\hat{\tilde{n}}_{\mathbf{i}}\hat{\tilde{n}}_{\mathbf{j}}}{4} \Big),
\end{aligned}
\label{eq:tJp}
\end{equation}
where $\hat{\tilde{n}}_{\mathbf{i}} = 2 - \hat{n}_{\mathbf{i}, \downarrow} - \hat{n}_{\mathbf{i}, \uparrow} = 0$ ($1$) for doublons (singly occupied sites).

We now map the doublon-doped $t$-$J$ model, Eq.~\eqref{eq:tJp}, to the hole-doped system, Eq.~\eqref{eq:tJh}, and describe both in the same Hilbert space. As a mere charge conjugation transformation $\hat{C} \hat{c}_{\mathbf{i}, \sigma} \hat{C}^{-1} =  \hat{c}_{\mathbf{i}, \sigma}^{\dagger}$, $\hat{C} \hat{c}_{\mathbf{i}, \sigma}^{\dagger} \hat{C}^{-1} = \hat{c}_{\mathbf{i}, \sigma}$, which maps particle creation to annihilation operators and vice versa, leads to phase and spin flips (see Appendix~\ref{sec:A1}), the charge conjugation operation can be redefined as
\begin{equation}
\begin{gathered}
    \hat{\mathcal{C}} \hat{c}_{\mathbf{i}, \sigma} \hat{\mathcal{C}}^{-1} = \eta_{\mathbf{i}} \text{sgn}(\bar{\sigma}) \hat{c}_{\mathbf{i}, \bar{\sigma}}^{\dagger} \\
    \hat{\mathcal{C}} \hat{c}_{\mathbf{i}, \sigma}^{\dagger} \hat{\mathcal{C}}^{-1} = \eta_{\mathbf{i}} \text{sgn}(\bar{\sigma}) \hat{c}_{\mathbf{i}, \bar{\sigma}}.
\end{gathered}
\label{eq:trafo}
\end{equation}
Here, the sign factor $\eta_{\mathbf{i}} = e^{i \boldsymbol{\pi} \cdot \mathbf{i}}$ with $\boldsymbol{\pi} = [\pi, \pi]$ is positive (negative) on sublattice A (B) on the square lattice and $\sigma = \uparrow, \downarrow$ is the spin state with $\bar{\sigma} = -\sigma$.

Applying the transformation to single particle states on a given site (see Appendix~\ref{sec:A1}) yields
\begin{equation}
    \begin{gathered}
        \hat{\mathcal{C}} \ket{0} = \hat{c}^{\dagger}_{\uparrow} \hat{c}^{\dagger}_{\downarrow} \ket{0} = \ket{\uparrow \downarrow} \\
        \hat{\mathcal{C}} \ket{\uparrow} = \hat{\mathcal{C}} \hat{c}^{\dagger}_{\uparrow} \hat{\mathcal{C}}^{-1} \hat{\mathcal{C}} \ket{0} = - \hat{c}^{\vphantom\dagger}_{\downarrow} \hat{c}^{\dagger}_{\uparrow} \hat{c}^{\dagger}_{\downarrow} \ket{0} = \ket{\uparrow} \phantom{,} \\
        \hat{\mathcal{C}} \ket{\downarrow} = \hat{\mathcal{C}} \hat{c}^{\dagger}_{\downarrow} \hat{\mathcal{C}}^{-1} \hat{\mathcal{C}} \ket{0} = \phantom{-} \hat{c}^{\vphantom\dagger}_{\uparrow} \hat{c}^{\dagger}_{\uparrow} \hat{c}^{\dagger}_{\downarrow} \ket{0} = \ket{\downarrow},
    \end{gathered}
\end{equation}
such that the singly occupied states map onto themselves, while doublons map to holes and vice versa.
\begin{figure*}
\centering
\includegraphics[width=0.67\textwidth]{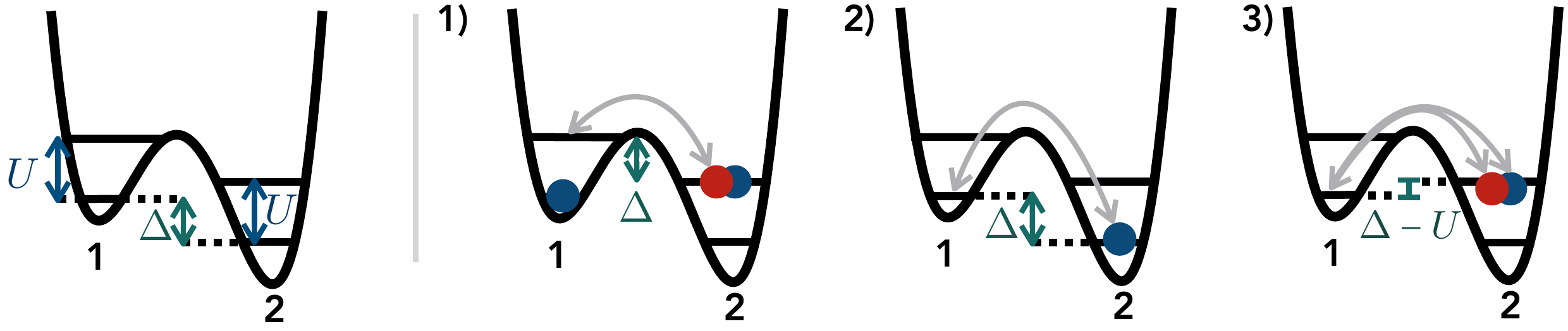}
\caption{\textbf{Interlayer interactions.} Left panel: illustration of the double well geometry with on-site interaction $U$ and energy offset $\Delta$; here, layer 1 (2) is the energetically upper (lower) layer. In the doublon-hole-doped setting, virtual processes lead to the appearance of nearest-neighbor interlayer interactions between particles. Contributions come from \textbf{1)} a particle in the upper ($\alpha = 1$) and a doublon in the lower ($\alpha = 2$) layer, \textbf{2)} a hole in the upper and a particle in the lower layer, and \textbf{3)} a hole in the upper and a doublon in the lower layer.}
\label{fig:V}
\end{figure*}

Transforming the relevant operators in the Hamiltonian Eq.~\eqref{eq:tJp} for nearest neighbor pairs $\braket{\mathbf{i}, \mathbf{j}}$ ultimately yields (see Appendix~\ref{sec:A1})
\begin{equation}
    \hat{\mathcal{C}} \hat{\tilde{\mathcal{H}}} \hat{\mathcal{C}}^{-1} = \hat{\mathcal{H}},
\end{equation}
such that the description of the doublon-doped system in the subspace of singly and doubly occupied sites is manifestly equivalent to a hole-doped description in the Hilbert space of singly occupied and empty sites (i.e., the $t$-$J$ Hamiltonian is particle-hole symmetric). Note that this is relying on the fact that the underlying lattice is bipartite (hence, including the additional 3-site term in the $t$-$J$ model does not change this result); non-bipartite lattices (e.g. when considering diagonal couplings $t'$~\cite{xu2023coexistence}) yield different signs in the hopping term after the charge-conjugation operation, and are hence not particle-hole symmetric. 

\subsection{Partial particle-hole mapping of the mixD $t$-$J$ model}
We now consider the hole-doped mixD bilayer $t_{\parallel}$-$J_{\perp}$-$J_{\parallel}$ model, which we propose to simulate,
\begin{equation}
    \hat{\mathcal{H}} = \sum_{\alpha=1,2} \hat{\mathcal{H}}_{\alpha} + \hat{\mathcal{H}}_{12}.
    \label{eq:SM_Hbl}
\end{equation}
Here, $\hat{\mathcal{H}}_{\alpha}$ denotes the Hamiltonian in layer $\alpha=1,2$ alone,
\begin{equation}
\begin{aligned}
    \hat{\mathcal{H}}_{\alpha} = -t_{\parallel} &\sum_{ \braket{\mathbf{i}, \mathbf{j}}, \sigma} \hat{\mathcal{P}} \big(\hat{c}_{\mathbf{i}, \sigma, \alpha}^{\dagger} \hat{c}_{\mathbf{j}, \sigma, \alpha}^{\vphantom\dagger} + \text{h.c.} \big) \hat{\mathcal{P}} \\ &+ J_{\parallel} \sum_{\braket{\mathbf{i}, \mathbf{j}}} \Big( \hat{\mathbf{S}}_{\mathbf{i},\alpha} \cdot \hat{\mathbf{S}}_{\mathbf{j},\alpha} - \frac{\hat{n}_{\mathbf{i},\alpha}\hat{n}_{\mathbf{j},\alpha}}{4} \Big) ,
\end{aligned}
\end{equation}
and $\hat{\mathcal{H}}_{12}$ is their interlayer Kondo-type coupling $\propto J_{\perp}$,
\begin{equation}
\begin{aligned}
    \hat{\mathcal{H}}_{12} = J_{\perp} \sum_{\mathbf{i}} \Big( \hat{\mathbf{S}}_{\mathbf{i},1} \cdot \hat{\mathbf{S}}_{\mathbf{i},2} - \frac{\hat{n}_{\mathbf{i},1}\hat{n}_{\mathbf{i},2}}{4} \Big).
\end{aligned}
\end{equation}
This model, with 50\% hole-doping, has been proposed to describe bilayer nickelate (LNO) high-$T_c$ superconductors under pressure~\cite{lu2023, oh2023type, qu2023}. In particular, it was shown that the model features (quasi) long-range $s$-wave pairing order, with expected critical temperatures of $T_c \sim J_{\perp}/2$ for $t_{\parallel}/J_{\perp} \sim 0.6$ in the 2D limit~\cite{schlömer2023superconductivity}.  

The Hamiltonian Eq.~\eqref{eq:SM_Hbl} can be simulated in bilayer optical lattices described by the on-site interaction $U$, intralayer (interlayer) tunnel couplings $\tilde{t}_{\parallel}$ ($\tilde{t}_{\perp}$), and a potential offset between the two layers $\Delta$~\cite{Hirthe2022, bourgund2023, chalopin2024optical}. When choosing $\tilde{t}_{\parallel}, \tilde{t}_{\perp} \ll \Delta < U$, the mixD setting in Eq.~\eqref{eq:SM_Hbl} is realized, with effective parameters $t_{\parallel} = \tilde{t}_{\parallel}$, $t_{\perp} = 0$, $J_{\parallel} = 4\tilde{t}_{\parallel}^2/U$ and $J_{\perp} = 2\tilde{t}_{\perp}^2/(U+\Delta) + 2\tilde{t}_{\perp}^2/(U-\Delta)$. 

Experimentally, we argue below that it is advantageous to simulate a closely related mixD bilayer model, with hole-doping in one and doublon-doping in the other layer. Using the notation from above, this model is described by the Hamiltonian
\begin{equation}
    \hat{\tilde{\mathcal{H}}} = \hat{\mathcal{H}}_1 + \hat{\tilde{\mathcal{H}}}_2 + \hat{\tilde{\mathcal{H}}}_{12}^{\prime}.
\end{equation}
Here, layer 1 (2) is hole (doublon) doped and the coupling $\hat{\tilde{\mathcal{H}}}^{\prime}_{12}$ contains additional interlayer density-density interactions arising from the mapping of a Hubbard model with a strong potential gradient between the layers, as derived in Refs.~\cite{Lange2023_1, Lange2023_2} and discussed in more detail next. The original motivation in Refs.~\cite{Lange2023_1, Lange2023_2} was to utilize the additional interlayer density-density interactions in the doublon-hole-doped model to tune the system through a crossover associated with a Feshbach resonance. Here we demonstrate another useful feature of this setting that readily allows to measure coherent pairing correlations. 

From now on, we assume that layer 1 is energetically offset from layer 2, i.e., holes (doublons) are doped in layer 1 (2). We shall denote the layer with larger (smaller) on-site energy as upper (lower) layer. On the left-hand side of Fig.~\ref{fig:V}, we schematically show a single double well with energy offset $\Delta$, where $\alpha = 1$ ($\alpha = 2$) correspond to the lower (upper) layer. We note that the situation of doublon (hole) doping the upper (lower) layer instead leads to the same conclusions as discussed in the following, however with a different sign of the additional density-density interactions. \\

\textbf{Density-density interactions. }
Before we apply a particle-hole mapping $\hat{\mathcal{C}}$ in the doublon-doped layer in order to relate $\hat{\tilde{\mathcal{H}}}$ to a mixD bilayer system with hole-doping ($\hat{\mathcal{H}}$), we describe the origin of the additional interactions in $\hat{\tilde{\mathcal{H}}}$, 
\begin{equation}
    \begin{aligned}
    \hat{\tilde{\mathcal{H}}}^{\prime}_{12} = J_{\perp} \sum_{\mathbf{i}} \hat{\mathbf{S}}_{\mathbf{i},1} &\cdot \hat{\mathbf{S}}_{\mathbf{i},2} \\ &+ \left( V-\frac{J_{\perp}}{4}\right) \sum_{\mathbf{i}} \hat{n}_{\mathbf{i},1}\hat{\tilde{n}}_{\mathbf{i},2}.
    \end{aligned}
\end{equation}
Virtual tunnel couplings between the doublon and hole-doped layer lead to the appearance of nearest neighbor (interlayer) interactions between dopants. The following contributions appear when doublon-doping the energetically lower layer ($\alpha = 2$) and hole-doping the upper layer ($\alpha = 1$)~\cite{Lange2023_1, Lange2023_2}, see Fig.~\ref{fig:V} (we note that we always imply $0<\Delta<U$):
\begin{enumerate}[{(1)}]
    \item $-\frac{\tilde{t}_{\perp}^2}{\Delta}\hat{n}_{\mathbf{i}, 1}(1-\hat{\tilde{n}}_{\mathbf{i}, 2})$ for a particle in the upper and a doublon in the lower layer,
    \item $-\frac{\tilde{t}_{\perp}^2}{\Delta}(1-\hat{n}_{\mathbf{i}, 1})\hat{\tilde{n}}_{\mathbf{i}, 2}$ for a hole in the upper and a particle in the lower layer,
    \item $-2\frac{\tilde{t}_{\perp}^2}{\Delta-U}(1-\hat{n}_{\mathbf{i}, 1})(1-\hat{\tilde{n}}_{\mathbf{i}, 2})$ for a hole in the upper and a doublon in the lower layer.
\end{enumerate}
Here, $\tilde{t}_{\perp}$ denotes the hopping between layers in the bilayer Fermi-Hubbard model with a gradient. Adding up the above contributions, we find an effective interlayer nearest neighbor interaction of the form
\begin{equation}
    V \sum_{\mathbf{i}} \hat{n}_{\mathbf{i},1} \hat{\tilde{n}}_{\mathbf{i}, 2},
    \label{eq:V}
\end{equation}
with $V = 2 \tilde{t}_{\perp}^2\left(\frac{1}{\Delta} + \frac{1}{U-\Delta}  \right)$. For $\Delta < U$, interactions are repulsive, $V>0$. As previously noted, the sign of $V$ changes when doublon (hole) doping the energetically upper (lower) layer instead: In this case, $V = -2 \tilde{t}_{\perp}^2\left(\frac{1}{\Delta} - \frac{1}{U+\Delta}  \right) < 0$ for all values of $\Delta, U >0$.

\begin{figure*}
\centering
\includegraphics[width=0.95\textwidth]{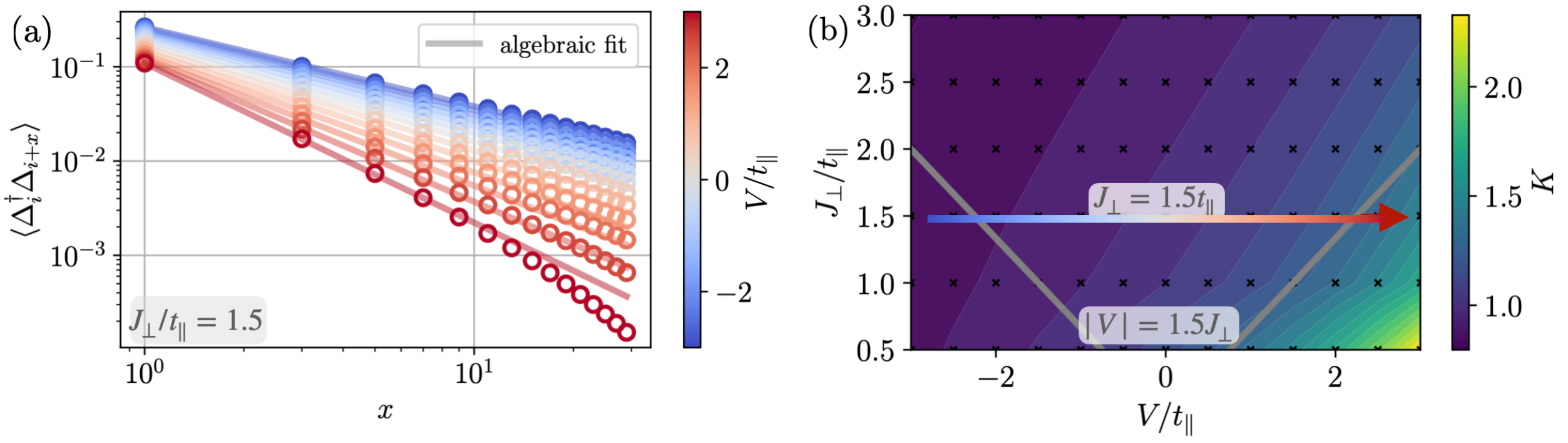}
\caption{\textbf{DMRG simulations.} (a) Coherent pairing correlations $\braket{\hat{\Delta}^{\dagger}_{i} \hat{\Delta}_{i+x}^{\vphantom \dagger}}$ in the ground state of a mixD ladder of length $L_x=60$ at hole doping $\delta=0.5$, for experimentally relevant couplings $J_\parallel/t_{\parallel}=0.5$ and $J_\perp/t_{\parallel} = 1.5$~\cite{Hirthe2022} and for various interaction strengths $V/t_{\parallel}$. Fits to algebraic decay functions are shown by solid lines; for illustrative reasons, only odd distances are shown. (b) Luttinger decay exponents $K$ as a function of $J_{\perp}/t_{\parallel}$ and $V/t_{\parallel}$. Contours denote extrapolated lines of constant $K$, black dots the points evaluated for the extrapolation. The arrow with the color gradient indicates the $V/t_{\parallel}$ scan in (a) from attractive (blue) to repulsive (red) interactions. Gray lines show typical interaction strengths for $\Delta=U/2$, $V=1.5J_\perp$, see Eq.~\eqref{eq:SM_Hbl_V}.}
\label{fig:DeltaiDeltaj}
\end{figure*}

Now we apply the partial particle-hole mapping $\hat{\mathcal{C}}_2 \equiv \mathbb{1}_1 \otimes \hat{\mathcal{C}}_2$, acting only on the doublon-doped layer 2, in order to obtain the effective Hamiltonian in the Hilbert space with hole dopants only. Since we are in mixed dimensions, no hopping terms $t_{\perp}$ exist in the effective $t$-$J$ description: Hence after applying the transformation $\hat{\mathcal{C}}$ from Eq.~\eqref{eq:trafo}, the doublon-doped layer changes to an equivalent hole-doped layer, $\hat{\mathcal{C}}_2 \hat{\tilde{\mathcal{H}}}_2 \hat{\mathcal{C}}_2^{-1} = \hat{\mathcal{H}}_2$. Moreover, using $\hat{\mathcal{C}}_2 \hat{\tilde{n}}_{\mathbf{i},2}\hat{\mathcal{C}}_{2}^{-1} = \hat{n}_{\mathbf{i},2}$ (see Appendix~\ref{sec:A1}), the coupling $\hat{\tilde{\mathcal{H}}}^{\prime}_{12}$ transforms and we obtain 
\begin{equation}
    \hat{\mathcal{C}}_2 \hat{\tilde{\mathcal{H}}} \hat{\mathcal{C}}_2^{-1} = \hat{\mathcal{H}}_1 + \hat{\mathcal{H}}_2 + \hat{\mathcal{H}}_{12} + V\sum_{\mathbf{i}}\hat{n}_{\mathbf{i}, 1} \hat{n}_{\mathbf{i}, 2}.
\end{equation}
Therefore, the physical implementation of the doublon-hole-doped mixD bilayer FH model after the particle-hole mapping corresponds to a fully hole-doped mixD bilayer system with interlayer hole-hole interactions, i.e., the simulated Hamiltonian reads
\begin{widetext}
\begin{equation}
\begin{aligned}
    \hat{\mathcal{H}} = -t_{\parallel} \sum_{ \braket{\mathbf{i}, \mathbf{j}}, \sigma, \alpha} \hat{\mathcal{P}} \big(\hat{c}_{\mathbf{i}, \sigma, \alpha}^{\dagger} \hat{c}_{\mathbf{j}, \sigma, \alpha}^{\vphantom\dagger} + \text{h.c.} \big)\hat{\mathcal{P}} + J_{\parallel} \sum_{\braket{\mathbf{i}, \mathbf{j}}, \alpha} &\Big( \hat{\mathbf{S}}_{\mathbf{i},\alpha} \cdot \hat{\mathbf{S}}_{\mathbf{j},\alpha} - \frac{\hat{n}_{\mathbf{i},\alpha}\hat{n}_{\mathbf{j},\alpha}}{4} \Big) \\ &+ J_{\perp} \sum_{\mathbf{i}} \hat{\mathbf{S}}_{\mathbf{i},1} \cdot \hat{\mathbf{S}}_{\mathbf{i},2} + \left(V - \frac{J_{\perp}}{4} \right)\sum_{\mathbf{i}} \hat{n}_{\mathbf{i},1}\hat{n}_{\mathbf{i},2}.
\end{aligned}
\label{eq:SM_Hbl_V}
\end{equation}
\end{widetext}
After the partial particle-hole transformation as specified above, the interactions in Eq.~\eqref{eq:V} correspond to tunable interlayer density-density interactions.
We note that in the case of hole or doublon-doping both layers in the physical implementation, virtual tunnel couplings between the two energetically offset layers merely lead to a constant energy shift; Hence, tunable density-density interactions $V$ are a particular feature of the doublon-hole-doped mixD bilayer system~\cite{Lange2023_1, Lange2023_2}.

\subsection{Phase-coherent pairing correlations}
Finally we explain why working with a doublon and a hole-doped layer provides a major experimental advantage. 
To this end, we note that resonant tunnel couplings between the two layers, if added again, lead to the appearance of terms like $\hat{c}^{\dagger}_{\mathbf{i}, \sigma, 1} \hat{c}^{\vphantom\dagger}_{\mathbf{i}, \sigma, 2}$. In the particle-hole transformed basis, these become $\hat{\mathcal{C}}_2 \hat{c}^{\dagger}_{\mathbf{i}, \sigma, 1} \hat{c}^{\vphantom\dagger}_{\mathbf{i}, \sigma, 2} \hat{\mathcal{C}}_2^{-1} \propto \hat{c}^{\dagger}_{\mathbf{i}, \sigma, 1} \hat{c}^{\dagger}_{\mathbf{i}, \bar{\sigma}, 2}$, which create and destroy pairs in the effective description of holes and singly occupied sites. In the mixD setting $\tilde{t}_{\parallel}, \tilde{t}_{\perp} \ll \Delta<U$, such pair-creation and annihilation terms do not appear. By explicitly tunnel coupling the two layers, the particle-hole transformed basis can be exploited to measure coherent pair-pair correlations in the mixD bilayer model without changing the total number of fermions, which we demonstrate in more detail in Sec.~\ref{sec:measurement}. In the following, we argue that the effective model Eq.~\eqref{eq:SM_Hbl_V} features long-range pairing order for a wide range of parameters in the ground state (as well as quasi long-range order at finite $T<T_c$). We then explain explicitly how pairing correlations can be measured for a simple double-well building block, before presenting state preparation and measurement protocols for mixD systems. \\

\textbf{Long-range superconducting order in the simulated model: numerical results. }
For a typical experimental value of $\Delta = U/2$, we obtain $\vert V\vert =1.5 J_{\perp}$. In the following, we show that this is a moderate repulsion which does not qualitatively change the physics and superconducting properties of the mixD model. We perform DMRG calculations~\cite{WhiteDMRG, Schollwoeck_DMRG, SchollwoeckDMRG2, hubig:_syten_toolk, hubig17:_symmet_protec_tensor_networ} of Eq.~\eqref{eq:SM_Hbl_V} on a ladder geometry, for varying $t_{\parallel}/J_{\perp}$, $V/J_{\perp}$ and hole doping $\delta=0.5$ in both layers. We explicitly exploit the $U(1)^{\alpha = 1} \times U(1)^{\alpha = 2}$ charge conservation symmetry in each leg.

\begin{figure*}
\centering
\includegraphics[width=0.74\textwidth]{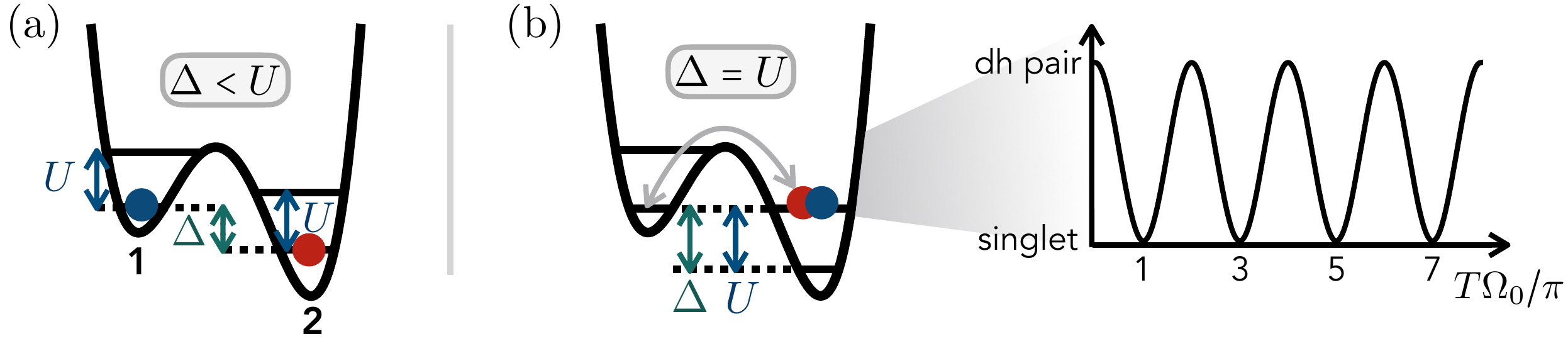}
\caption{\textbf{Double well Rabi oscillations.} (a) A single double well building block with energy shift $\Delta < U$. The ground state with one particle per site is a singlet state. (b) When tuning $\Delta = U$ to resonance, coherent Rabi oscillations (in time $T$) between singlets and doublon-hole pairs are induced, with holes (doublons) situated in the energetically upper (lower) site, labeled site 1 (2).}
\label{fig:dw}
\end{figure*}

Fig.~\ref{fig:DeltaiDeltaj}~(a) shows pair-pair correlations $\braket{\hat{\Delta}^{\dagger}_{i} \hat{\Delta}_{i+x}^{\vphantom \dagger}}$ for a ladder of length $L_x = 60$, with $J_{\perp}/t_{\parallel} = 1.5$ as approximately realized in Ref.~\cite{Hirthe2022} and for varying interlayer density-density interactions $V/t_{\parallel} = -3 \dots 3$. In a gapless 1D system, pairing correlations decay algebraically, $\braket{\hat{\Delta}^{\dagger}_{i} \hat{\Delta}_{i+x}^{\vphantom \dagger}} \sim x^{-K}$, with $K$ the Luttinger parameter. Corresponding fits for $i=10$ and $x=1,\dots,42$ are shown by solid lines in Fig.~\ref{fig:DeltaiDeltaj}~(a). For a broad range of interaction strengths $V/t_{\parallel}$, clear algebraic signals are found. Only for strong repulsive interactions $V >V_c\approx  2.5 t_{\parallel} \approx 1.7 J_\perp$, we observe the onset of an exponential decay for $\braket{\hat{\Delta}^{\dagger}_{i} \hat{\Delta}_{i+x}^{\vphantom \dagger}}$ and large distances $x$. This is in agreement with previous observations of a pair charge gap opening at commensurate doping $\delta=0.5$ and intermediate repulsion $V$ \cite{Lange2023_2}, see also Appendix~\ref{appendix:Gaps}. 

Fig.~\ref{fig:DeltaiDeltaj}~(b) presents $K$ as a function of both $V/t_{\parallel}$ and $J_{\perp}/t_{\parallel}$. Again, for most considered values of $V/t_{\parallel}$ and $J_{\perp}/t_{\parallel}$ we find Luttinger parameters $K\approx 1$. Only for large $V>V_c \approx 3.0 t_{\parallel} = 2.0 J_\perp$ and $J_\perp/t_\parallel<1.5$, $K$ becomes significantly larger before the onset of an exponential decay.
The Kondo coupling $J_\perp/t_\parallel=1.5$ and the corresponding $\vert V\vert=1.5J_\perp$ [indicated by the gray lines in Fig. \ref{fig:DeltaiDeltaj}~(b)] realized in the experiment \cite{Hirthe2022} are deep in the $K\approx 1$ regime for attractive $V$ (i.e. doublon-doping the energetically upper layer), and for repulsive interactions (i.e. doublon-doping the energetically lower layer) $K\approx 1.5$. Note that it is possible to tune $J_\perp/t_\parallel$ and $V/t_\parallel$, e.g. via the potential offset $\Delta$, to the regime with smaller $K$, i.e. longer-ranged correlations.  

We note that at finite temperature, correlations decay exponentially in ladder systems due to their one-dimensional nature~\cite{giamarchi2003quantum}. Nevertheless, in the same spirit as early observations of antiferromagnetic order~\cite{Greif2013, Hart2015, Boll2016, Mazurenko_AFM, Hilker2017}, measurable finite-range correlations are expected in regimes realistically accessible to ultracold atom experiments. In particular, in~\cite{schlömer2023superconductivity} it was shown that binding energies in ladder systems exceed $E_b/J_{\perp} > 0.5$ for all values of $t_{\parallel}/J_{\perp}$. Therefore, for temperatures $T/E_b \lesssim 1$, we expect sizable pairing correlations in ladder systems. In contrast, in the 2D bilayer limit, we expect algebraic correlations up to relatively high critical BKT temperatures even in the presence of interlayer density repulsions. Indeed, in the perturbative limit of strong interlayer couplings, the system maps to a 2D hard-core bosonic system in which critical temperatures of $T_c \sim J_{\perp}/2$ have been estimated~\cite{schlömer2023superconductivity}. \\

\textbf{A single double well. }
As an experimental building block, consider a single double well with energy offset $\Delta < U$, loaded with two fermions ($N_{\uparrow} = N_{\downarrow} = 1$). The ground state corresponds to a spin singlet, $\ket{\text{s}} = \frac{1}{\sqrt{2}} (\ket{\uparrow}_1 \ket{\downarrow}_2 - \ket{\downarrow}_1 \ket{\uparrow}_2)$, see Fig.~\ref{fig:dw}~(a). When tuning the tunneling transition between doublon-hole pairs $\ket{\text{dh}} = \ket{0}_1 \ket{\uparrow \downarrow}_2$ and singlets $\ket{\text{s}}$ into resonance, i.e. setting $\Delta = U$, hopping transitions are induced, 
\begin{equation}
    \begin{aligned}
        \sum_\sigma \hat{c}^{\dagger}_{\sigma, 2} \hat{c}^{\vphantom\dagger}_{\sigma, 1} &\ket{\text{s}} =\frac{1}{\sqrt{2}} \Big( \hat{c}^{\dagger}_{\uparrow, 2} \hat{c}^{\vphantom\dagger}_{\uparrow, 1} \ket{\uparrow}_1 \ket{\downarrow}_2 \\ &\underbrace{ - \hat{c}^{\dagger}_{\downarrow, 2} \hat{c}^{\vphantom\dagger}_{\downarrow, 1} \ket{\downarrow}_1 \ket{\uparrow}_2}_{ = + \ket{0}_1 \ket{\uparrow \downarrow}_2} \Big) \propto \ket{\text{dh}}. 
    \end{aligned}
    \label{eq:hopping1}
\end{equation}
Similarly, 
\begin{equation}
    \begin{aligned}
        \sum_\sigma \hat{c}^{\dagger}_{\sigma, 1} \hat{c}^{\vphantom\dagger}_{\sigma, 2} &\ket{\text{dh}} = \frac{1}{\sqrt{2}}\Big( \hat{c}^{\dagger}_{\uparrow, 1} \hat{c}^{\vphantom\dagger}_{\uparrow, 2} \ket{0}_1 \ket{\uparrow \downarrow}_2 \\ &+ \underbrace{\hat{c}^{\dagger}_{\downarrow, 1} \hat{c}^{\vphantom\dagger}_{\downarrow, 2} \ket{0}_1 \ket{\uparrow \downarrow}_2}_{= - \ket{\downarrow}_1 \ket{\uparrow}_2}\Big) \propto \ket{\text{s}}. 
    \end{aligned}
    \label{eq:hopping2}
\end{equation}
This gives rise to Rabi oscillations between singlets and doublon-hole pairs, Fig.~\ref{fig:dw}~(b). Double well Rabi oscillations of single particles have been demonstrated with high fidelity in Refs.~\cite{Murmann2015, Yang2020, impertro2023local, chalopin2024optical}.

When mapping the doublon-hole-doped system to the fully hole-doped mixD $t$-$J$ basis via the partial particle-hole transformation, we now see that the tunneling operations in Eqs.~\eqref{eq:hopping1},\eqref{eq:hopping2} formally correspond to spin-singlet creation and annihilation operators, 
\begin{equation}
\begin{aligned}
        \begin{split}
        \hat{\mathcal{C}}_2( \hat{c}^{\dagger}_{\uparrow, 1} \hat{c}^{\vphantom\dagger}_{\uparrow, 2} + &\hat{c}^{\dagger}_{\downarrow, 1} \hat{c}^{\vphantom\dagger}_{\downarrow,2} ) \hat{\mathcal{C}}_2^{-1} \\ &= \underbrace{\hat{c}^{\dagger}_{\downarrow, 1} \hat{c}^{\dagger}_{\uparrow, 2} - \hat{c}^{\dagger}_{\uparrow, 1} \hat{c}^{\dagger}_{\downarrow,2}}_{\propto \hat{\Delta}^{\dagger}}
        \end{split} \\
        \begin{split}
        \hat{\mathcal{C}}_2( \hat{c}^{\dagger}_{\uparrow, 2} \hat{c}^{\vphantom\dagger}_{\uparrow, 1} + &\hat{c}^{\dagger}_{\downarrow, 2} \hat{c}^{\vphantom\dagger}_{\downarrow, 1} ) \hat{\mathcal{C}}_2^{-1}  \\&=  \underbrace{ \hat{c}^{\vphantom\dagger}_{\downarrow, 1} \hat{c}^{\vphantom\dagger}_{\uparrow, 2} - \hat{c}^{\vphantom\dagger}_{\uparrow, 1} \hat{c}^{\vphantom\dagger}_{\downarrow, 2}}_{\propto \hat{\Delta}}.
        \end{split} 
        \end{aligned}
\end{equation}

\begin{figure*}
\centering
\includegraphics[width=0.65\textwidth]{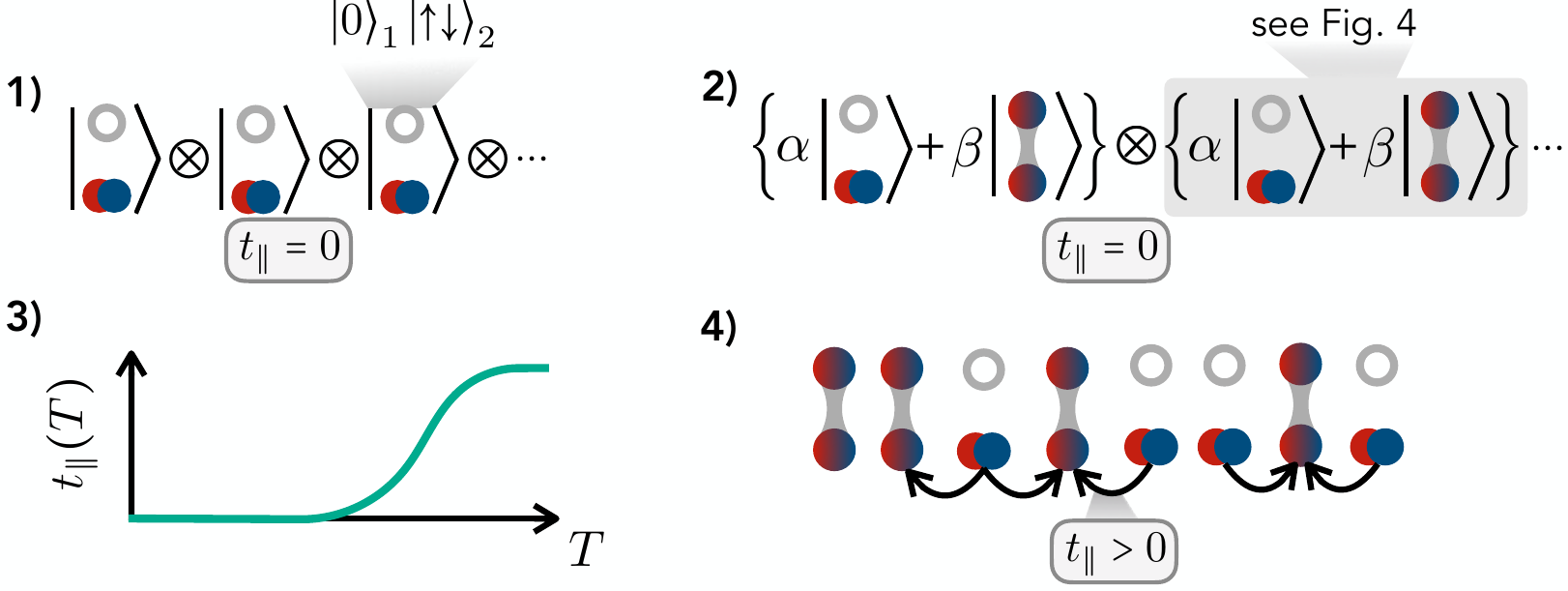}
\caption{\textbf{State preparation scheme.} State preparation for simulating the mixD $t$-$J$ model on a ladder. \textbf{1)} A band insulating state is prepared in deep double wells with $\Delta > U$. \textbf{2)} Tuning $\Delta = U$ for a given amount of time, singlets are created with a certain average density $|\beta|^2$ (where still $t_{\parallel} = 0$), cf. Fig.~\ref{fig:dw}. Afterwards, $\Delta$ is tuned to $\Delta < U$. \textbf{3)} Hopping between double wells is adiabatically turned on by ramping down the lattice potentials, which enables the simulation of the mixD bilayer model on a ladder geometry in step \textbf{4)}.}
\label{fig:prep}
\end{figure*}

Therefore, the Rabi oscillations between doublon-hole and singlet states in the physical system map onto coherent oscillations between a rung-singlet and a paired state of two holes after applying the partial particle-hole transformation $\hat{\mathcal{C}}_2$ on the second (doublon-doped) layer. Thus, measurements of singlet-to-doublon-hole oscillations in the physical system provide a direct measurement of coherent pair creation in the corresponding $\hat{\mathcal{C}}_2$-transformed system that we ultimately propose to investigate. We finally note that tuning $\Delta$ of the optical superlattice to resonance has the advantage of high control, while avoiding lattice shaking induced transitions which cause heating.

\subsection{State preparation scheme}
Based on coupled double-wells as building blocks, in the following we present a state preparation scheme for simulating the mixD system. We note that ultimately, the simulation of the full 2D bilayer model Eq.~\eqref{eq:SM_Hbl_V} is desirable. For this purpose, bilayer capabilities of existing quantum gas microscope experiments~\cite{Preiss2015, Gall2021Bilayer} used e.g. for spin-resolved imaging~\cite{Gross_bilayer, Koepsell_bilayer2020} can be utilized. Offsetting the two layers by an energy $\Delta$ then implements the mixD model on a 2D square lattice geometry. However, as a first step, we argue that implementing mixD ladders as in Ref.~\cite{Hirthe2022} already constitutes a valuable setup that is readily available to measure coherent pairing correlations in 1D systems. For this purpose, the state preparation consists of the following steps, summarized in Fig.~\ref{fig:prep}: 

\begin{enumerate}[{(1)}]
\item Loading doublon-hole pairs into multiple (separate) double wells with strong potential gradients $\Delta > U$ realizes the product state
\begin{equation}
    \bigotimes_{\mathbf{i}} \ket{0, \uparrow \downarrow}_{\mathbf{i}},
\end{equation}
where $\mathbf{i}$ denotes the index of the double wells. Lattice potentials are sufficiently deep, such that $t_{\parallel} = 0$. This can be achieved with low entropy starting from a band insulating state~\cite{Chiu2018_bi} in the lower layer.
\item By globally tuning the optical superlattice to resonance ($\Delta = U$), singlets are coherently created. The fraction of doublon-hole pairs compared to singlets can be controlled depending on the time $T_{\text{res}}$ for which resonant tunneling is switched on. After $T_{\text{res}}$, $\Delta$ is further ramped down to $\Delta < U$. This allows for the preparation of e.g. a 50:50 mixture (50\% doublon/hole doping) as present in LNO. 
\item When adiabatically ramping down the lattice depth, in-plane hopping $t_{\parallel}$ is induced, simulating the mixD bilayer system Eq.~\eqref{eq:SM_Hbl_V}. When considering a ladder geometry, adiabatic ramping of $t_{\parallel}$ guarantees that a low-temperature Luther-Emery liquid of phase-coherent pairs~\cite{Lange2023_1, Lange2023_2} is realized.
\end{enumerate}
This scheme can also be directly applied to a mixD bilayer constituted by two coupled 2D layers. Note that the phase-coherent creation of pairs in step (2) readily realizes the product state $\bigotimes_{\mathbf{i}} \left(\alpha \ket{0, \uparrow\downarrow}_{\mathbf{i}}+\beta \ket{s}_{\mathbf{i}}\right)$, which has long-range pairing correlations and is hence adiabatically connected to low-energy states of the mixD $t$-$J$ model. We expect that after sufficient thermalization times, the system reaches an equilibrium steady-state whose correlation functions correspond to those of the mixD system, Eq.~\eqref{eq:SM_Hbl_V}. This suggests high fidelities reachable in step (3). 

The accuracy of the global $\pi/2$ pulse ultimately sets the doping value of the simulated mixD bilayer model. We note that away from $50\%$ doping, the system remains superconducting, with only slight renormalization of pair-pair correlations~\cite{schlömer2023superconductivity}. Thus, we expect that the measurement output is stable against infidelities of the state preparation scheme, e.g. due to local fluctuations of the Hamiltonian parameters. 

\subsection{Measurement protocol}
\label{sec:measurement}
After the adiabatic state preparation of the doublon-hole-doped mixD system, coherent pair-pair correlations can be measured. In particular, in the fully hole-doped target system that we propose to realize in a particle-hole transformed basis, we would like to directly measure correlations $\braket{\hat{\Delta}_{\mathbf{i}}^{\dagger} \hat{\Delta}_{\mathbf{j}}}$, where $\hat{\Delta}_{\mathbf{i}}^{\dagger} = \frac{1}{\sqrt{2}} (\hat{c}^{\dagger}_{\mathbf{i}, \uparrow, 1} \hat{c}^{\dagger}_{\mathbf{i}, \downarrow, 2} - \hat{c}^{\dagger}_{\mathbf{i}, \downarrow, 1} \hat{c}^{\dagger}_{\mathbf{i}, \uparrow,2})$ creates an interlayer singlet on site $\mathbf{i}$. This requires measuring $\hat{\mathcal{C}}_2 \hat{\Delta}_{\mathbf{i}}^{\dagger} \hat{\Delta}_{\mathbf{j}} \hat{\mathcal{C}}_2^{-1}$ in the physical system with doublon (hole) doping in the lower (upper) layer.

In the following, we first describe the protocol in the limit of strong Kondo-couplings $J_{\perp} \gg t_{\parallel}, J_{\parallel}$, where a mapping to an effective spin-$1/2$ system reveals how coherent pairing correlations can be accessed through a basis rotation in the subspace of singlets and doublon-hole pairs. Subsequently, we extend the discussion to situations away from the perturbative regime. \\

\begin{figure*}
\centering
\includegraphics[width=0.92\textwidth]{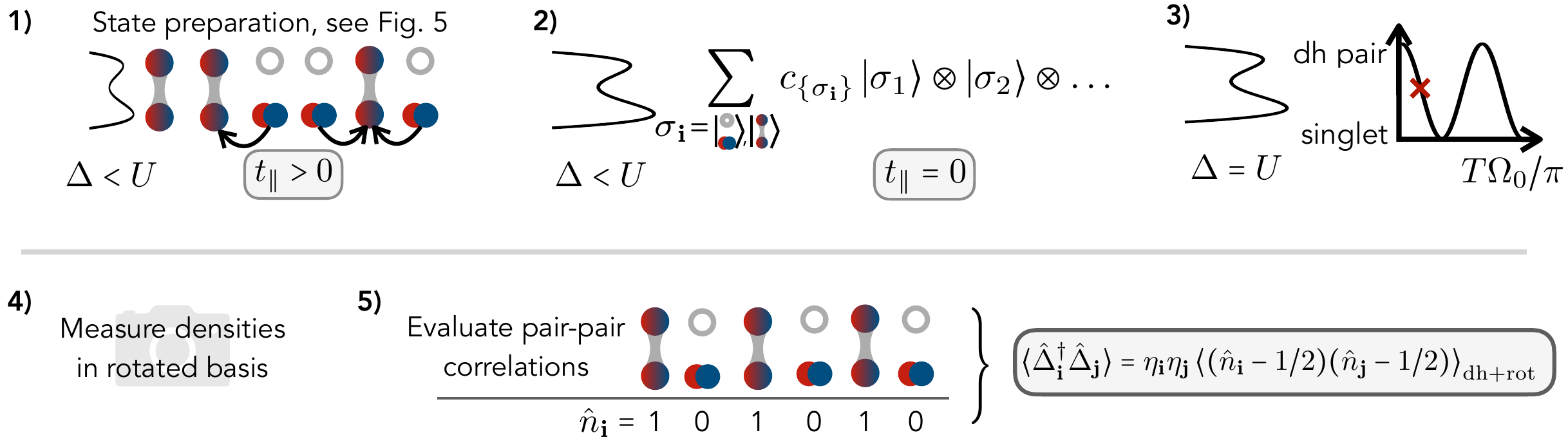}
\caption{\textbf{Measurement protocol.} After state preparation [cf. Fig.~\ref{fig:prep}] in step \textbf{1)}, lattice potentials are ramped up to freeze out $t_{\parallel} = 0$, \textbf{2)}. A $\pi/2$ basis rotation by tuning $U=\Delta$ \textbf{3)} and consecutively measuring densities of singlets and doublon-hole pairs \textbf{4)} then allows for evaluation of pair-pair correlations in step \textbf{5)}.}
\label{fig:measurement}
\end{figure*}

\textbf{Perturbative limit. }
In the case of strong Kondo-couplings $J_{\perp} \gg t_{\parallel}, J_{\parallel}$, fermions pair into tightly bound interlayer singlets with associated binding energies $J_{\perp}$. In this limit, the low-energy Hilbert space of Eq.~\eqref{eq:SM_Hbl} is spanned by chargon-chargon pairs (i.e., holes on site $\mathbf{i}$ in both layers, $\ket{0}_{\mathbf{i}} = \ket{0}_{\mathbf{i},1} \ket{0}_{\mathbf{i},2}$), and rung-singlets, $\ket{1}_{\mathbf{i}} = \hat{b}^{\dagger}_{\mathbf{i}} \ket{0}_{\mathbf{i}}$. Here, we have defined the (hard-core) bosonic operator $\hat{b}^{\dagger}$ that creates an interlayer spin singlet on site $\mathbf{i}$, $\hat{b}^{\dagger}_{\mathbf{i}} \ket{0}_{\mathbf{i}} = \frac{1}{\sqrt{2}}\left( \hat{c}^{\dagger}_{\mathbf{i},\uparrow, 1} \hat{c}^{\dagger}_{\mathbf{i},\downarrow,2} - \hat{c}^{\dagger}_{\mathbf{i},\downarrow,1} \hat{c}^{\dagger}_{\mathbf{i},\uparrow,2} \right) \ket{0}_{\textbf{i}}$, with $\hat{b}^{\dagger}_{\mathbf{i}} \hat{b}_{\mathbf{i}} \leq 1$ due to the hard-core constraint. As derived in Ref.~\cite{Bohrdt2020} (see also Refs.~\cite{Lange2023_1, Lange2023_2, schlömer2023superconductivity}), by considering second order perturbative processes where interlayer singlets are virtually destroyed, the mixD bilayer model Eq.~\eqref{eq:SM_Hbl} maps to a hard-core bosonic system with nearest neighbor interactions on a single-layer square lattice,
\begin{widetext}
\begin{equation}
\begin{aligned}
    \hat{\mathcal{H}}_{\text{HCB}} = &-\frac{K}{2} \sum_{ \braket{\mathbf{i}, \mathbf{j}}} \hat{\mathcal{P}} \big(\hat{b}_{\mathbf{i}}^{\dagger} \hat{b}_{\mathbf{j}}^{\vphantom\dagger} + \text{h.c.} \big)\hat{\mathcal{P}} - J_{\perp} \sum_{\mathbf{i}} \hat{b}_{\mathbf{i}}^{\dagger} \hat{b}_{\mathbf{i}}^{\vphantom\dagger}  &+ K \sum_{\braket{\mathbf{i}, \mathbf{j}}} \left( \delta \hat{b}_{\mathbf{i}}^{\dagger} \hat{b}_{\mathbf{i}}^{\vphantom\dagger} \hat{b}_{\mathbf{j}}^{\dagger} \hat{b}_{\mathbf{j}}^{\vphantom\dagger} - \frac{\hat{b}_{\mathbf{i}}^{\dagger} \hat{b}_{\mathbf{i}}^{\vphantom\dagger}}{2} - \frac{\hat{b}_{\mathbf{j}}^{\dagger} \hat{b}_{\mathbf{j}}^{\vphantom\dagger}}{2}  \right),
\end{aligned}
\label{eq:Hhcb}
\end{equation}
\end{widetext}
where $K = 4 t_{\parallel}^2/J_{\perp}$ and $\delta = 1-J_{\parallel}/2K$. 

For what follows, it is useful to associate the chargon-chargon pair and interlayer singlet with two spin states of an effective XXZ model described by spin-1/2 operators $\hat{J}^{\mu}_{\mathbf{i}}$, $\mu = x,y,z$ (see e.g. Ref.~\cite{sachdev_2023_qpm}), 
\begin{equation}
    \begin{gathered}
        \hat{J}_{\mathbf{i}}^{+} = \eta_{\mathbf{i}} \hat{b}^{\dagger}_{\mathbf{i}}, \quad  
         \hat{J}_{\mathbf{i}}^{-} = \eta_{\mathbf{i}} \hat{b}_{\mathbf{i}} \\
         \hat{J}^z_{\mathbf{i}} = \hat{b}^{\dagger}_{\mathbf{i}} \hat{b}_{\mathbf{i}} - 1/2,
    \end{gathered}
    \label{eq:trafo_2}
\end{equation}
where again $\eta_{\mathbf{i}}$ is 1 (-1) on the A (B) sublattice. The hard-core bosonic model Eq.~\eqref{eq:Hhcb} then maps to a 2D XXZ spin model,
\begin{equation}
\begin{aligned}
    \hat{\mathcal{H}}_{\text{XXZ}} = K \sum_{\braket{\mathbf{i}, \mathbf{j}}} \left( \hat{J}^x_{\mathbf{i}} \hat{J}^x_{\mathbf{j}} + \hat{J}^y_{\mathbf{i}} \hat{J}^y_{\mathbf{j}} + \delta \hat{J}^z_{\mathbf{i}} \hat{J}^z_{\mathbf{j}} \right).
\end{aligned}
\label{eq:XXZ}
\end{equation}
We note that in Eq.~\eqref{eq:XXZ}, constant terms that arise have been dropped. The magnetization of the spin model relates to the filling $n$ of the bilayer model through $m = n - 1/2$. In particular, we note that for $J_{\parallel} = 0$, Eq.~\eqref{eq:XXZ} reduces to the 2D Heisenberg model with an emerging $\rm{SU(2)}$ symmetry.

Pair-pair correlations in the mixD $t$-$J$ model map to in-plane spin-spin correlations of the XXZ Hamiltonian, $\braket{\hat{\Delta}_{\mathbf{i}}^{\dagger} \hat{\Delta}_{\mathbf{j}}} \rightarrow \eta_{\mathbf{i}} \eta_{\mathbf{j}} \braket{\hat{J}^+_{\mathbf{i}} \hat{J}^{-}_{\mathbf{j}}}$. 
These can be accessed through a basis rotation in the subspace of singlets and hole pairs, in analogy to measurements of in-plane (off-diagonal) spin-spin correlations in the FH model~\cite{Brown2017}, $\braket{\hat{J}^+_{\mathbf{i}} \hat{J}^{-}_{\mathbf{j}}} = 2 \braket{\hat{J}^x_{\mathbf{i}} \hat{J}^{x}_{\mathbf{j}}} =  2 \braket{\hat{J}^z_{\mathbf{i}} \hat{J}^{z}_{\mathbf{j}}}_{\text{rot}}$. In particular, we propose the following measurement scheme, summarized in Fig.~\ref{fig:measurement}:
\begin{enumerate}[{(1)}]
    \item State preparation, see Fig.~\ref{fig:prep}.
    \item Ramp up lattice depth to freeze in-plane degrees of freedom, $t_{\parallel} = 0$.
    \item Rotate basis by a global $\pi/2$ tunneling pulse, see Fig.~\ref{fig:dw}.
    \item Measure densities (which corresponds to a measurement of the mapped spin $\hat{\mathbf{J}}$ in the XXZ model in the $z$-basis after the rotation (3), i.e., either a doublon-hole pair or singlet is measured at each site $\mathbf{i}$).
    \item Using Eq.~\eqref{eq:trafo_2}, evaluate pair-pair correlations via 
\begin{equation}
    \begin{aligned}
    \braket{\hat{\Delta}_{\mathbf{i}}^{\dagger} \hat{\Delta}_{\mathbf{j}}} = 2\eta_{\mathbf{i}} \eta_{\mathbf{j}} \Big\langle &(\hat{n}_{\mathbf{i}} - 1/2) \\ & \times (\hat{n}_{\mathbf{j}} - 1/2)\Big\rangle_{\text{dh+rot}}
    \end{aligned}
    \label{eq:measurement}
\end{equation}
Here, $\braket{\circ}_{\text{dh+rot}}$ denotes measurement of the doublon-hole-doped system in the rotated basis and $\hat{n}_{\mathbf{i}} = 1$ $(0)$ for a measured singlet (doublon-hole pair) on site $\mathbf{i}$ in the rotated basis.
\end{enumerate}
\textbf{Away from the perturbative limit. }
\begin{figure*}
\centering
\includegraphics[width=0.65\textwidth]{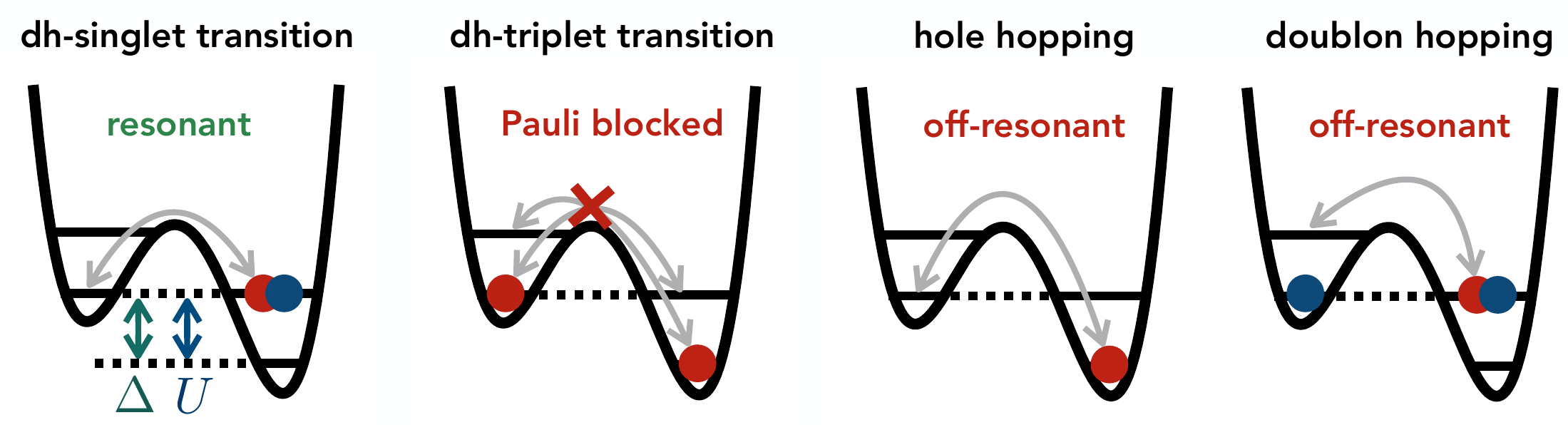}
\caption{\textbf{Allowed and blocked transitions.} Applying a $\pi/2$ pulse by tuning $\Delta = U$ leads to the following effect depending on the local configuration: Transitions between local rung-singlets and doublon-hole pairs are resonant. Meanwhile, transitions between local rung-triplets and doublon-hole pairs are Pauli-blocked and interlayer hopping of single holes and doublons is off-resonant.}
\label{fig:transitions}
\end{figure*}
When tuning the system away from the perturbative limit, hopping processes can break interlayer singlets. The basis states of a single rung hence not only include singlets and doublon-hole pairs, but consist of $d=9$ states $\{ \ket{0}_1 \ket{\uparrow \downarrow}_2, \ket{\sigma}_1 \ket{\uparrow \downarrow}_2, \ket{0}_1 \ket{\sigma}_2, \ket{\text{s}}, \ket{\text{t}_{1,2,3}}\}$, where $\ket{\text{s}}, \ket{\text{t}_{1,2,3}}$ denote the rung singlet and three triplet states, respectively. A global $\pi/2$ tunneling pulse as described above realizes a basis rotation only in the subspace of the states $\{ \ket{\text{dh}}, \ket{\text{s}}\}$, as all other transitions are either Pauli blocked or off-resonant as shown in Fig.~\ref{fig:transitions}. 

When taking snapshots in the rotated basis, only contributions from singlets and doublon-hole pairs contribute to Eq.~\eqref{eq:measurement}. The two triplet states $\ket{\uparrow}_1 \ket{\uparrow}_2$, $\ket{\downarrow}_1 \ket{\downarrow}_2$ as well as the states $\ket{\sigma}_1 \ket{\uparrow \downarrow}_2$, $\ket{0}_1 \ket{\sigma}_2$ are trivially identified in spin-resolved snapshots and have zero contribution to Eq.~\eqref{eq:measurement}. Spin-resolved measurements in the mixD setting have been demonstrated on ladder geometries in Ref.~\cite{Hirthe2022}, which can be extended to 2D bilayers with current technologies. In order to also distinguish the triplet $\ket{\uparrow}_1 \ket{\downarrow}_2 + \ket{\downarrow}_1 \ket{\uparrow}_2$ from the singlet $\ket{\uparrow}_1 \ket{\downarrow}_2 - \ket{\downarrow}_1 \ket{\uparrow}_2$, we propose to adiabatically apply an $\hat{S}^z$ magnetic field gradient along the rungs before the final measurement, during which (see Appendix~\ref{appendix:sto_adia})
\begin{equation}
    \begin{gathered}
        \frac{1}{\sqrt{2}} \left( \ket{\uparrow}_1 \ket{\downarrow}_2 - \ket{\downarrow}_1 \ket{\uparrow}_2 \right) \rightarrow \ket{\uparrow}_1 \ket{\downarrow}_2 \\
        \frac{1}{\sqrt{2}} \left( \ket{\uparrow}_1 \ket{\downarrow}_2 + \ket{\downarrow}_1 \ket{\uparrow}_2 \right) \rightarrow \ket{\downarrow}_1 \ket{\uparrow}_2     
    \end{gathered}
\end{equation}
up to overall phases, while all other configurations remain unaffected. We note that this is in contrast to singlet-triplet oscillations, where a quench under a magnetic field gradient leads to coherent oscillations between singlet and triplet states~\cite{Trotzky, Trotzky2010, Greif2013, Murmann2015}.

During the measurement scheme, infidelities of global basis rotations and adiabatic ramping of the magnetic field gradient lead to quantitative deviations of the measurement signal to the true pair-pair correlations. However, we stress that signals consistent with finite values away from zero constitute evidence for the existence of pair-pair correlations in the system. Therefore we argue that, while ultimately obtaining quantitatively accurate results is desirable, the qualitative interpretation of non-zero signals influenced by infidelities remains the same. This paves the way towards a systematic exploration of superconducting phases in inherently repulsively interacting fermionic systems with state-of-the-art ultracold atom simulators.

\section{Measuring pairing correlations: 2D Fermi-Hubbard model}
\label{sec:FH_pairing}
We have demonstrated that in the mixD setting, coherent pair-creation and annihilation processes can be naturally implemented through tunneling transitions between doublon- and hole-doped layers. We now show that related ideas allow for the measurement of pairing correlations in the 2D (single-layer) FH model, through a partial particle-hole transformation that maps the attractive to the repulsive FH model. In particular, we show that local control enables access to spin-singlet pairing correlations between both horizontal and vertical bonds, $P^{\alpha, \beta}_{\mathbf{i}, \mathbf{j}} = \langle \hat{\Delta}^{\dagger}_{\mathbf{i},\alpha} \hat{\Delta}_{\mathbf{j},\beta}\rangle$, where $\hat{\Delta}^{\dagger}_{\mathbf{i}, \mu} = \frac{1}{\sqrt{2}} \left(\hat{c}^{\dagger}_{\mathbf{i}, \uparrow} \hat{c}^{\dagger}_{\mathbf{i}+\hat{\mathbf{e}}_{\mu}, \downarrow} - \hat{c}^{\dagger}_{\mathbf{i}, \downarrow} \hat{c}^{\dagger}_{\mathbf{i}+\hat{\mathbf{e}}_{\mu}, \uparrow}\right)$ with $\mathbf{\hat{\mathbf{e}}}_{\mu}$ the unit lattice vectors along $\mu = x,y$. For $s$-wave pairing, all combinations of $\alpha, \beta$ yield the same sign in the correlator, whereas a $d$-wave pairing structure features different signs of $P^{xy}$ compared to $P^{xx}$, $P^{yy}$. This way, our scheme allows for a direct observation of the sign and nodal structure of the pair-pair correlations in the 2D doped FH model, which can be identified as a superconducting order parameter.

We note that pairing correlations between parallel bonds, i.e. $\alpha = \beta$, are easier to access experimentally compared to the case $\alpha \neq \beta$, where in the latter two optical superlattices with different orientations need to be realized, e.g. by using local addressability. Nevertheless, independent of the pairing symmetry, $P_{\mathbf{i}, \mathbf{j}}^{\alpha = \beta} \neq 0$ at large distances $|\mathbf{i}-\mathbf{j}| \gg 1$ if the system features superconducting order. Hence, working with a superlattice with one orientation and fixing $\alpha = \beta$, pairing correlations (including $d$-wave) can be detected in the 2D FH model (though the symmetry can not be uniquely specified in this case). 


Our proposed measurement scheme relies on a particle-hole symmetry of the 2D FH model, which in turn requires the underlying lattice to be bipartite. State-of-the-art numerical calculations suggest the absence of long-range superconducting order in the ground state of the plain-vanilla 2D Fermi-Hubbard model on the square lattice (i.e. without next-nearest neighbor terms $t'$)~\cite{Qin_absence_SC}. While finite values of $t'$ are believed to stabilize superconductivity~\cite{xu2023coexistence} in the ground state, at typical temperatures of quantum simulation platforms correlations are expected to decay exponentially in both the plain-vanilla and finite $t'$ FH model. Therefore, independent on the (short- or long-range) pairing structure of the ground state, at first only exponentially decaying pair-pair correlations can be accessed with currently available technologies. We argue that it is these short-range correlations (including their symmetry) that can be probed with our measurement scheme, which, in turn, constitutes first valuable steps towards simulating and probing superconductivity in 2D FH-type models. Furthermore, we stress that even in this regime many open questions can be addressed,
including the microscopic nature of the formation of pairs and the relation between the
pseudogap and superconductivity.

\subsection{Partial particle-hole mapping of the 2D Fermi-Hubbard model}
Consider the repulsive ($U>0$) 2D FH model on the square lattice at finite doping which we would ultimately like to simulate,
\begin{equation}
    \begin{aligned}
    \hat{\mathcal{H}}(t,U) = -t \sum_{\braket{\mathbf{i}, \mathbf{j}}, \sigma} \big( \hat{c}^{\dagger}_{\mathbf{i}, \sigma} &\hat{c}_{\mathbf{j}, \sigma} + \text{h.c.} \big) \\&+ U \sum_{\mathbf{i}} \hat{n}_{\mathbf{i},\uparrow} \hat{n}_{\mathbf{i},\downarrow}.
    \end{aligned}
\end{equation}
On a bipartite lattice, a partial particle-hole mapping of one spin-species 
\begin{equation}
    \hat{\mathcal{C}} \hat{c}_{\mathbf{i}, \sigma} \hat{\mathcal{C}}^{-1} = \begin{cases}
        \hat{c}_{\mathbf{i}, \uparrow} \quad \quad \text{for } \sigma = \uparrow  \\
        \eta_{\mathbf{i}} \hat{c}_{\mathbf{i}, \downarrow}^{\dagger} \quad \,  \text{for } \sigma = \downarrow
    \end{cases}
\label{eq:trafo_ar}
\end{equation}
transforms (up to an overall constant) the repulsive to the attractive FH model~\cite{Ho2009} (see also Refs.~\cite{Gall2020, Hartke2023}), i.e.,
\begin{equation}
    \hat{\mathcal{C}}^{-1} \hat{\mathcal{H}}(t, U) \hat{\mathcal{C}} = \hat{\mathcal{H}}(t, -U) + U \sum_{\mathbf{i}} n_{\mathbf{i}, \uparrow}.
\end{equation}
Note that from Eq.~\eqref{eq:trafo_ar} follows that the vacuum state of the repulsive model $\ket{0}$ transforms as
$\hat{\mathcal{C}} \ket{0} = \prod_{\mathbf{i}} \hat{c}_{\mathbf{i}, \downarrow}^{\dagger} \ket{0}$. Thus, a hole-pair $\ket{0}_1 \ket{0}_2 = \ket{0,0}$ on neighboring sites $1,2$ in the repulsive model corresponds to a spin-down pair $\hat{\mathcal{C}}\ket{0}_1\ket{0}_2 = \ket{\downarrow}_1 \ket{\downarrow}_2 = \ket{\downarrow, \downarrow}$ in the attractive model. Accordingly, hole doping the repulsive system translates to a finite magnetization on the attractive side, $\hat{\mathcal{C}}^{-1}\hat{N}_h\hat{\mathcal{C}} = \hat{N}_{\downarrow} - \hat{N}_{\uparrow}$, where $N_{h,\uparrow,\downarrow}$ denote the total number of holes, up- and down-spins, respectively. 

\begin{figure*}
\centering
\includegraphics[width=0.95\textwidth]{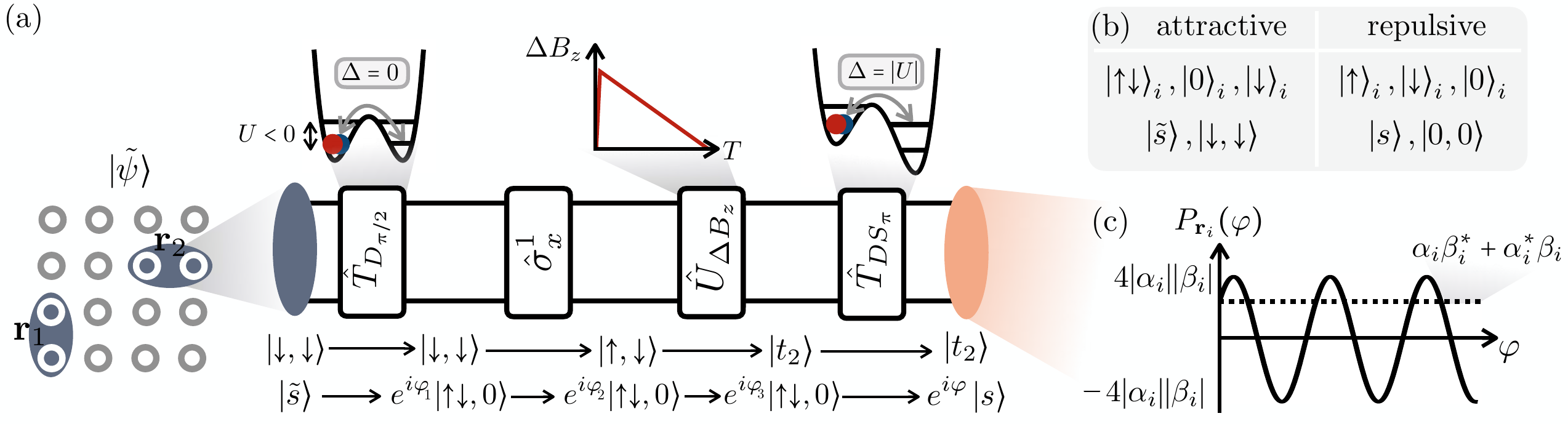}
\caption{\textbf{Measuring pairing correlations in the plain-vanilla FH model.} (a) Circuit that allows to measure coherent pair-pair correlations in the repulsive FH model between bonds $\mathbf{r}_1$, $\mathbf{r}_2$ through implementation of the attractive FH model. The mapping from the attractive to the repulsive side is summarized in (b). The shown sequence in (a) consists of a combination of local tunneling and spin gates. The former are realized by locally reducing the lattice depth and tuning the superlattice offset $\Delta$. Adiabatic tuning of a magnetic field gradient $\Delta B_z$ allows to map $\ket{\downarrow,\downarrow}$ [corresponding to the hole-pair state in the repulsive model, see (b)] to a spin-triplet, whereas $\ket{\tilde{s}}$ [corresponding to $\ket{s}$ in the repulsive model, see (b)] is transformed to a spin-singlet. After each step, relative phases between the two input states are picked up, with an overall phase shift of $\varphi$ by the end of the sequence. Spin-resolved measurements after applying the unitary allow for the evaluation of pairing correlations $P_{\mathbf{i}, \mathbf{j}}^{\alpha, \beta}$ in the repulsive FH model. (c) The overall phase difference $\varphi$ between the final output states gives rise to oscillations of measurement probabilities $P_{\mathbf{r}_i}$, from which $\langle\big(\hat{\Delta}^\dagger_{\mathbf{r}_1} + \hat{\Delta}_{\mathbf{r}_1}\big)\big(\hat{\Delta}^\dagger_{\mathbf{r}_2} + \hat{\Delta}_{\mathbf{r}_2}\big)\rangle = (\alpha_1\beta_1^* + \alpha_1^* \beta_1)(\alpha_2\beta_2^* + \alpha_2^* \beta_2)$ can be reconstructed.}
\label{fig:measurement_FH}
\end{figure*}

Furthermore, the spin-singlet state $\ket{s}$ in the repulsive Hamiltonian maps to
\begin{equation}
\begin{aligned}
    \ket{\tilde{s}} = \hat{\mathcal{C}} \ket{s} = \frac{1}{\sqrt{2}} \left( \ket{\uparrow \downarrow}_1 \ket{0}_2 + \ket{0}_1 \ket{\uparrow \downarrow}_2 \right) 
\end{aligned}
\end{equation}
in the attractive model, see also Fig.~\ref{fig:measurement_FH}~(b) for a summary of the transformation.

In the following,  we show how the application of local unitary gates which map $\ket{t_1} = \ket{\downarrow}_1 \ket{\downarrow}_2$ and $\ket{\tilde{s}}$ to the unmagnetized spin-triplet and singlet state, respectively, allows to measure the pairing operator. In particular, spin-resolved measurement statistics in the $z$-basis after applying the gates give access to pair-pair correlations in the repulsive FH model.

\subsection{Measurement protocol}
Starting from a low-temperature state of the attractive FH model ($U<0$) in an optical lattice, lattice depths are ramped up globally to freeze any dynamics. The goal is to measure the pair correlator in the repulsive FH model $\langle \hat{\Delta}^\dagger_{\mathbf{r}_1}\hat{\Delta}_{\mathbf{r}_2}\rangle$ on bonds  $\mathbf{r}_1=(\mathbf{i},\mathbf{i}+\hat{\mathbf{e}}_\mu)$ and $\mathbf{r}_2=(\mathbf{j},\mathbf{j}+\hat{\mathbf{e}}_{\mu^\prime})$, each of them aligned along the unit vectors $\hat{\mathbf{e}}_{\mu^{(\prime)}}$. 

To understand how this can be achieved we find it convenient to decompose the quantum many-body wave function into the following (orthogonal) contributions: $(i)$ States with two holes at bond $\mathbf{r}_i$ in the repulsive model, corresponding to $\ket{\downarrow,\downarrow}_i$ in the attractive model, $(ii)$ the states with a singlet pair at bond $\mathbf{r}_i$, corresponding to $\ket{\tilde{s}}_i=\frac{1}{\sqrt{2}} \left( \ket{\uparrow\downarrow,0}_i + \ket{0,\uparrow\downarrow}_i\right) $ in the attractive model, and $(iii)$ orthogonal contributions $\ket{\phi}$. Notably, applications of $\hat{\Delta}_{\mathbf{r}_i}$ only act within the subspace spanned by $(i)$ and $(ii)$.

Therefore, a general pure state on bond $\mathbf{r}_i$ in the repulsive model reads
\begin{align}
    \ket{\psi}_i =\alpha_i \ket{0,0}_i+\beta_i\ket{s}_i+\ket{\phi}.
\end{align}
Correspondingly, we get in the attractive model 
\begin{align}
    \ket{\tilde{\psi}}_i = \alpha_i \ket{\downarrow,\downarrow}_i+\beta_i\ket{\tilde{s}}_i+\ket{\tilde{\phi}}.
\end{align}
To motivate how our measurement scheme works, we assume a product state $\ket{\Psi} = \ket{\psi}_1\ket{\psi}_2$ that explicitly breaks the $\mathrm{U(1)}$ particle conservation symmetry next. However, in Appendix~\ref{appendix:2DHubbard} we provide a proof for arbitrary correlated states. The scheme still works in the latter case since we only employ unitary operations acting independently on the different bonds $\mathbf{r}_i$.

Expectation values of the pairing fields in the repulsive model yield
\begin{equation}
\begin{aligned}
    &_i\langle \psi \vert \hat{\Delta}^\dagger_{\mathbf{r}_i}\vert \psi\rangle_i = \alpha_i\beta_i^{*}, \\
    &_i\langle \psi \vert \hat{\Delta}_{\mathbf{r}_i}\vert \psi\rangle_i = \alpha_i^{*}\beta_i,
\end{aligned}
\end{equation}
and therefore 
\begin{equation}
\begin{aligned}
\langle\Psi|\big(\hat{\Delta}^\dagger_{\mathbf{r}_1} &+ \hat{\Delta}_{\mathbf{r}_1}\big)\big(\hat{\Delta}^\dagger_{\mathbf{r}_2} + \hat{\Delta}_{\mathbf{r}_2}\big)|\Psi\rangle \\ &= (\alpha_1\beta_1^* + \alpha_1^* \beta_1)(\alpha_2\beta_2^* + \alpha_2^* \beta_2). 
\label{eq:Deltas}
\end{aligned}
\end{equation}
Hence, by measuring certain products of $\alpha_1^{(*)}, \alpha_2^{(*)}, \beta_1^{(*)}, \beta_2^{(*)}$ in the attractive model, pairing correlations in the repulsive model can be accessed. 

In the following, we introduce a measurement circuit that realizes a unitary $\hat{U}^\varphi =\prod_i \hat{U}_{\mathbf{r}_i}^\varphi$, which allows to extract such products $\alpha_i^{(*)}\beta_i^{(*)}$ from Fock-basis snapshots of the transformed state. We design the measurement protocol such that the action of this unitary on the state realizes a Ramsey-type interferometer, 
\begin{equation}
\begin{aligned}
    \hat{U}_{\mathbf{r}_i}^\varphi \ket{\tilde{\psi}}_i= \Big(&\gamma^{+,\varphi}_i\ket{\uparrow,\downarrow}_i+\gamma^{-,\varphi}_i\ket{\downarrow,\uparrow}_i \Big) +\ket{\phi},
\end{aligned}
\end{equation}
with $\gamma^{\pm,\varphi}_i=\frac{1}{\sqrt{2}}(\alpha_i\pm \mathrm{e}^{i\varphi}\beta_i)$, see Fig. \ref{fig:measurement_FH}~(a). 

Before we discuss the experimental realization of $\hat{U}^\varphi$ in Sec.~\ref{sec:U}, we explain how it gives access to the pair correlations Eq.~\eqref{eq:Deltas}:
Consider the probability to measure $\ket{\uparrow,\downarrow}_{\mathbf{r}_i}$ after applying $\hat{U}_{\mathbf{r}_i}^\varphi$,
\begin{equation}
\begin{aligned}
    P^{\uparrow\downarrow}_{\mathbf{r}_i}(\varphi)&= \vert _{i}\langle \uparrow,\downarrow\vert \hat{U}_{\mathbf{r}_i}^\varphi\vert \tilde{\psi}\rangle_i\vert^2 \\&= \frac{1}{2}\vert \alpha_i + \mathrm{e}^{i\varphi}\beta_i\vert^2,
\end{aligned}
\end{equation}
and $P^{\downarrow\uparrow}_{\mathbf{r}_i}$ defined in analogy.
The difference between these probabilities,
\begin{equation}
\begin{aligned}
    P_{\mathbf{r}_i}(\varphi):&=(P^{\uparrow\downarrow}_{\mathbf{r}_i}-P^{\downarrow\uparrow}_{\mathbf{r}_i})(\varphi)\\&= 2\mathrm{e}^{-i\varphi}\left( \alpha_i \beta_i^{*}+\mathrm{e}^{2i\varphi}\alpha_i ^{*}\beta_i\right),
\end{aligned}
\end{equation}
evaluated at each of the two bonds $\mathbf{r}_i$ hence gives, via Born's rule, access to $\langle \hat{\Delta}^\dagger_{\mathbf{r}_i} + \hat{\Delta}_{\mathbf{r}_i}\rangle$ from the measurement statistics of $\downarrow\uparrow$ and $\uparrow\downarrow$ configurations. 

The phase dependency $P_{\mathbf{r}_i}(\varphi)$ gives rise to Ramsey fringes, i.e. oscillations between $\pm 4|\alpha_i||\beta_i|$, with $P_{\mathbf{r}_i}(\varphi = 0) = \alpha_i \beta_i^* + \alpha_i^* \beta_i = \langle \hat{\Delta}^\dagger_{\mathbf{r}_i} + \hat{\Delta}_{\mathbf{r}_i}\rangle $, see Fig.~\ref{fig:measurement_FH}~(b). Correspondingly, by computing $P_{\mathbf{r}_1} P_{\mathbf{r}_2} = \langle\Psi|\big(\hat{\Delta}^\dagger_{\mathbf{r}_1} + \hat{\Delta}_{\mathbf{r}_1}\big)\big(\hat{\Delta}^\dagger_{\mathbf{r}_2} + \hat{\Delta}_{\mathbf{r}_2}\big)|\Psi\rangle$, pair-pair correlations can be computed.

We note that when the $\mathrm{U(1)}$ particle conservation symmetry is spontaneously broken, the order parameter averaged over many independent experimental realizations vanishes, $_i\langle \psi \vert \hat{\Delta}^{(\dagger)}_{\mathbf{r}_i}\vert \psi\rangle_i = 0$, and correspondingly $P_{\mathbf{r}_i} = 0$. Nevertheless, for general correlated states, the product of the probabilities $P_{\mathbf{r}_i} P_{\mathbf{r}_j}$ (corresponding to pair-pair correlations $\braket{\hat{\Delta}_{\mathbf{r}_i}^{\dagger} \hat{\Delta}_{\mathbf{r}_j} + \text{h.c.}}$) is finite at large distances $|\mathbf{r}_i - \mathbf{r}_j|$ for states with superconducting order\footnote{For general correlated states, pairing correlations are given by $\braket{\hat{\Delta}^{\dagger}_{\mathbf{r
}_1} \hat{\Delta}_{\mathbf{r
}_2}} = \text{tr}\left(\hat{\rho}_{12}^{sh} \hat{\Delta}^{\dagger}_{\mathbf{r
}_1} \hat{\Delta}_{\mathbf{r
}_2}\right)$, where $\hat{\rho}_{12}^{sh}$ is the reduced density matrix in the subspace of hole-pairs and singlets on bonds $\mathbf{r}_1, \mathbf{r}_2$ (this is justified as $\hat{\Delta}^{(\dagger)}_{\mathbf{r
}_i}$ only acts on this subspace). A similar analysis to the case of a product state shows that determining the probabilities of measuring $\ket{\uparrow,\downarrow}_i$ and $\ket{\downarrow, \uparrow}_i$ independently gives access to $\text{tr}\left[\hat{\rho}_{12}^{sh} \big(\hat{\Delta}^\dagger_{\mathbf{r}_1} \hat{\Delta}_{\mathbf{r}_2} + \hat{\Delta}^\dagger_{\mathbf{r}_2}\hat{\Delta}_{\mathbf{r}_1}\big)\right]$, see Appendix~\ref{appendix:2DHubbard}.}.

In particular, in this case $P_{\mathbf{r}_i} P_{\mathbf{r}_j}$ only depends on the overall phase difference picked up during the measurement scheme on the two bonds. When applying a control phase $\varphi_{\text{control}}$ to one of the two bonds, Ramsey fringes as a function of $\varphi_{\text{control}}$ can be observed in $P_{\mathbf{r}_i} P_{\mathbf{r}_j} (\varphi_{\text{control}})$. Though pairing correlations can be directly accessed by setting $\varphi_{\text{control}} = 0$, observation of the Ramsey fringes constitutes a valuable experimental verification of the coherence of the interferometer. 

\subsection{The measurement circuit \label{sec:U}}
The measurement protocol that we propose consists of a sequence of unitaries that is applied locally on all bonds $\mathbf{r}_i$ independently. We focus on the states that contribute to the pair correlations, namely $\ket{\downarrow,\downarrow}_{\mathbf{r}_i}$ and $\ket{\tilde{s}}_{\mathbf{r}_i}$, that correspond to hole and singlet pairs on bond $\mathbf{r}_i$ in the repulsive model. The general idea of the measurement protocol is to apply a unitary $\hat{U}_{\mathbf{r}_i}^{\varphi}$ that transforms the states  $\ket{\downarrow,\downarrow}_i$ and $\ket{\tilde{s}}_i$ to spin-singlets and triplets (up to a controlled overall phase difference $\varphi$), respectively, from which pair-pair correlations can be accessed by taking spin-resolved measurements in the $z$-basis. Note that all other contributions are orthogonal and remain orthorgonal during the sequence due to the unitarity of $\hat{U}_{\mathbf{r}_i}^{\varphi}$. The gate sequence that realizes $\hat{U}_{\mathbf{r}_i}^{\varphi}$ consists of the following steps, summarized in Fig.~\ref{fig:measurement_FH}~(a).
\begin{enumerate}[{(1)}]
    \item First, by locally turning on tunneling (i.e. by reducing the lattice depth of double wells on sites connected by bonds $\mathbf{r}_{i}$), Rabi oscillations between the states $\ket{\uparrow\downarrow,0}$ and $\ket{0,\uparrow\downarrow}$ are induced with effective doublon tunneling rate $4t^2/|U|$. Starting from $\ket{\tilde{s}}$, a $\pi/2$ tunneling pulse $\hat{T}_{D_{\pi/2}}$ transforms the state to $\ket{\uparrow\downarrow,0}$. Meanwhile, tunneling transitions of $\ket{\downarrow,\downarrow}$ are Pauli blocked, i.e. the state stays invariant (up to an overall dynamical phase) under $\hat{T}_{D_{\pi/2}}$. 
    \item Spin-flip pulses $\sigma^1_x$ on the first sites of the respective double wells (i.e. on sites $\bf{i}$, $\bf{j}$) transforms $\ket{\downarrow,\downarrow}\rightarrow \ket{\uparrow, \downarrow}$, while $\ket{\uparrow\downarrow,0}$ remains unchanged. 
    \item By applying an adiabatic magnetic field gradient ramp along $z$, the transformation $\ket{\uparrow,\downarrow} \rightarrow \ket{t_2} = \frac{1}{\sqrt{2}}(\ket{\uparrow,\downarrow}+\ket{\downarrow,\uparrow})$ is performed (see Appendix~\ref{appendix:UBz}); Again, up to an overall phase, the doublon-hole state $\ket{\uparrow\downarrow,0}$ remains invariant.  
    \item Lastly, resonant oscillations between doublons and singlets ($\Delta = |U|$) allow to realize a $\pi$-pulse between the doublon-hole and singlet state, $\hat{T}_{DS_{\pi}}\ket{\uparrow\downarrow,0} \propto \ket{s}$. The triplet state $t_2$, on the other hand, is Pauli-blocked from corresponding tunnel transitions. 
\end{enumerate}
During this sequence, a relative phase $\varphi$ between the final singlet and triplet states is picked up. By applying a potential gradient after the first tunneling unitary, this overall relative phase $\varphi$ can be varied and the Ramsey fringes observed. This, in turn, ultimately allows to measure pair-pair correlations $\langle \hat{\Delta}^\dagger_{\mathbf{r}_1}\hat{\Delta}_{\mathbf{r}_2} + \hat{\Delta}_{\mathbf{r}_1}\hat{\Delta}^\dagger_{\mathbf{r}_2}\rangle$. In particular, by varying $\mu, \mu'$ for a given pair of sites $\mathbf{i}, \mathbf{j}$, $d$-wave pairing correlations can be measured by evaluating  
\begin{equation}
\begin{aligned}
    \langle \hat{\Delta}_{\mathbf{i},x}^{\dagger} \hat{\Delta}_{\mathbf{j},x} &+  \hat{\Delta}_{\mathbf{i},y}^{\dagger} \hat{\Delta}_{\mathbf{j},y} \\&- \hat{\Delta}_{\mathbf{i},x}^{\dagger} \hat{\Delta}_{\mathbf{j},y}
    -\hat{\Delta}_{\mathbf{i},y}^{\dagger} \hat{\Delta}_{\mathbf{j},x} + \text{h.c.}\rangle.
\end{aligned}
\end{equation}   

\section{Measuring dopant properties in the 2D $t$-$J$ model}
\label{sec:tJ}
The mixD setting does not only allow to investigate bilayer systems such as nickelate superconductors, but can also be mapped to a 2D single-layer $t$-$J$ model in some limits. In a similar spirit to doped carrier mean-field formulations of the $t$-$J$ model \cite{Ribeiro2005,Ribeiro2006,Pepino_2008}, we use an enlarged Hilbert space comprising one mixD rung per site, and map the low energy states in the large $J_\perp,U\gg t_\parallel$ limit of the mixD bilayer with a half-filled upper layer to the 2D $t$-$J$ model states. 

Compared to direct 2D cold atom realizations of the FH model~\cite{Bohrdt2020,Gross2017}, our proposal features two main advantages: Firstly, the holes/dopants of the effective $t$-$J$ model are represented by particles in an auxiliary layer, enabling to directly measure hole properties such as the momentum resolved hole density $\braket{\hat{n}_h(\mathbf{k})}$. This is particularly intriguing since $\braket{\hat{n}_h(\mathbf{k})}$ does not depend on the spectral weight at momentum $\mathbf{k}$, and can hence give access to regions of the Brillouin zone that cannot be investigated with ARPES, such as the backside of the Fermi arcs \cite{Shen2005}. In contrast, the direct implementation of the 2D $t$-$J$ model is limited to momentum resolved particle densities. Furthermore, the hopping and superexchange amplitudes of the effective $t$-$J$ model can be tuned independently from each other as they originate from different layers of the mixD bilayer. In contrast, in 2D realizations of the  $t$-$J$ model without the mapping that will be introduced below, the effective parameters are fixed by the relation $J=4t^2/U$. 

Below, we introduce the mapping from the physical, mixD bilayer to a 2D $t$-$J$ model and the required parameter regimes before turning to the discussion of the measurement protocol for determining the momentum resolved dopant density of the effective 2D $t$-$J$ model.

\begin{figure*}
\centering
\includegraphics[width=0.82\textwidth]{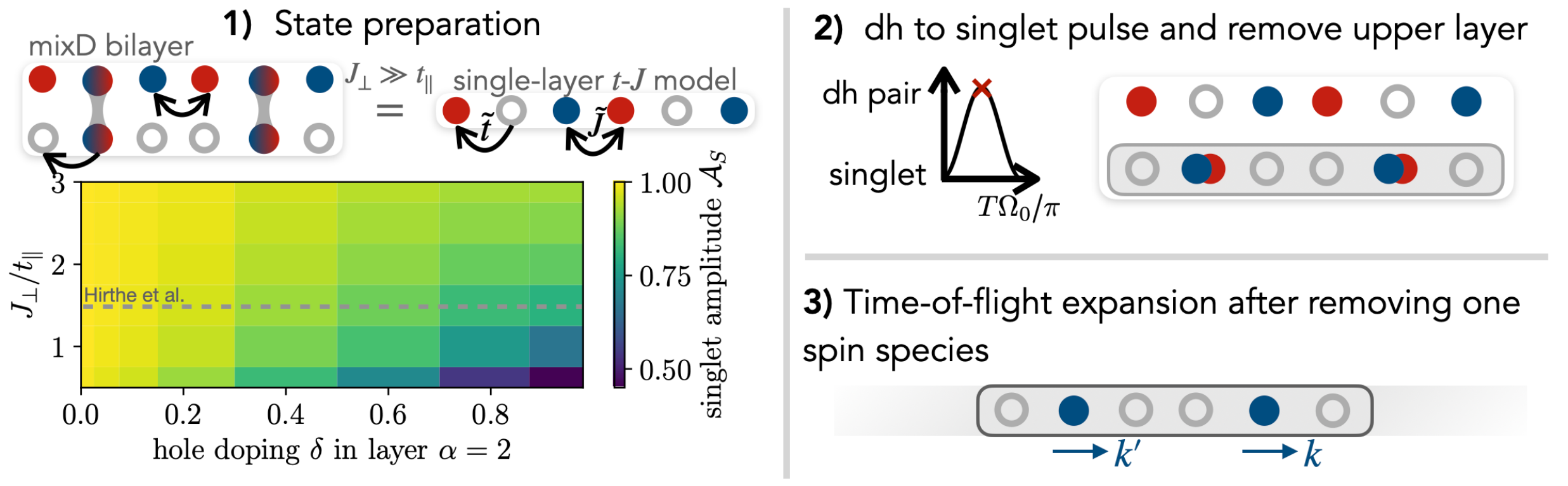}
\caption{\textbf{Measurement protocol of momentum resolved density.} \textbf{1)} State preparation: For large $J_\perp/t_\parallel$, singlets form on the rungs of the mixD layer and allow to map the system to a 2D single-layer $t$-$J$ model. We also show numerical results for the rung-singlet probability $\mathcal{A}_S$ (see Eq. \eqref{eq:As}) obtained for a ladder of length $L=40$ and different hole dopings of the lower leg $\delta$ (i.e. hole doping $1-\delta$ of the effective $t$-$J$ model). We use $J_\perp/t_\parallel$ and $J_\parallel/t_\parallel=0.07$ as used in Hirthe et al. \cite{Hirthe2022}. After the state preparation, a $\pi$ basis rotation is applied \textbf{2)} which maps all singlets to dh pairs with doublons in the lower $\alpha=2$ layer and holes in the upper $\alpha=1$ layer. When removing one spin species from the lower layer and ramping down the lattice depth \textbf{3)}, the remaining indistuingishable particles -- one per dopant -- are free and their momentum can be measured using a time-of-flight protocol. }
\label{fig:measurement_tJ}
\end{figure*}

\subsection{Hole-doping the one layer}
In order to map the mixD bilayer to a 2D $t$-$J$ model, we consider the mixD bilayer with layer-dependent parameters $U_\alpha$ and $t_\parallel^\alpha$ in the large $U_\alpha, J_\perp \gg t_\parallel^\alpha$ limit. In this limit, the large on-site repulsion $U_\alpha$ in both layers leads to maximally singly occupied sites in the whole system, and the large interlayer Kondo coupling $J_\perp $ to a ground state at half-filling consisting of singlets at each rung between the layers. 

The filling in each layer can be controlled individually. Here, we consider a half-filled (energetically) upper layer ($\n_{\mathbf{i},\alpha=1}=1$) and $\n_{\mathbf{i},\alpha=2}\leq 1$. In this case, the low energy states are either the (bosonic) rung singlets $\Bd_{\mathbf{i}}$ or singly occupied rungs with an empty site in layer $\alpha=2$, represented by fermionic operators $\hat{\Tilde{c}}^\dagger_{\mathbf{i},\sigma}=\Cd_{\mathbf{i},\sigma,1} \B_{\mathbf{i}}$, see Appendix \ref{appendix:2DtJ}.  We would like to point out that $\hat{\Tilde{c}}^\dagger_{\mathbf{i},\sigma}$ annihilates a rung singlet before creating a particle in layer $\alpha=1$, i.e. a particle of layer $\alpha=2$ is removed.  Furthermore, only the lower layer $\alpha=2$ contributes to the in-plane hopping terms since the particles in the half-filled upper layer are blocked due to the single particle constraint. Vice versa, all spin interactions arise from particles of the upper layer $\alpha=1$ at singly occupied rungs since the lower layer features either empty sites or particles that are part of rung singlets, both without any contribution to the spin exchange. This is taken into account by defining the spin operator in the low energy subspace,
\begin{align}
    \hat{\Tilde{\mathbf{S}}}_{\mathbf{i}} = \hat{\Tilde{c}}^\dagger_{\mathbf{i},\mu}\frac{\boldsymbol{\sigma}_{\mu \nu}}{2} \hat{\Tilde{c}}_{\mathbf{i}\nu} =  \hat{{\mathbf{S}}}_{\mathbf{i},1}(1-\Bd_\mathbf{i}\B_\mathbf{i}),
\end{align}
which only act on the singly occupied rungs and not on the singlets.

\subsection{Mapping to a 2D (single-layer) $t$-$J$ model}
With these considerations, we arrive at an effective single-layer model,
\begin{widetext}
\begin{equation}
\begin{aligned}
    \Ham_\mathrm{eff} = \mathcal{P}\left[ -\Tilde{t}\sum_{\langle \mathbf{i},\mathbf{j}\rangle}\hat{\Tilde{c}}^\dagger_{\mathbf{i}\sigma}\hat{\Tilde{c}}_{\mathbf{j}\sigma} + \hc \right]\mathcal{P}
    +  \Tilde{J} \sum_{\langle \mathbf{i},\mathbf{j} \rangle} \left[  \Tilde{\mathbf{S}}_\mathbf{i} \cdot \Tilde{\mathbf{S}}_\mathbf{j} - \frac{\gamma}{4}\hat{\Tilde{n}}_{i} \hat{\Tilde{n}}_{j}\right]+\frac{J_\parallel^{\alpha=2}}{2}\sum_{ \mathbf{i}}\Hat{\Tilde{n}}_\mathbf{i},
    \label{eq:Bilayer_to_tJ}
\end{aligned}
\end{equation}
\end{widetext}
with  $\Tilde{t}=t_\parallel^{\alpha=2}/2$, $\Tilde{J}=J_\parallel^{\alpha=1}$ and $\gamma=\frac{J_\parallel^{\alpha=2}}{\Tilde{J}}$, see Fig. \ref{fig:measurement_tJ} \textbf{1)} and Appendix \ref{appendix:2DtJ}. This model corresponds to a $t$-$J$ model with an additional chemical potential. Note that this model is formulated w.r.t. to a vacuum of only rung singlets of the actual mixD bilayer: creating holes in the actual model w.r.t. this vacuum corresponds to creating particles in the effective $t$-$J$ model and vice versa. Additionally, a three-site term leads to a next-nearest neighbor hopping of dopants when considering the corresponding three-site term of the $t-J$ layer $\alpha=2$, see Appendix \ref{appendix:2DtJ}.

Compared to a direct realization of the 2D $t$-$J$ layer, two advantages become apparent in Eq. \eqref{eq:Bilayer_to_tJ}: $(i)$ In principle, $\Tilde{t}$ and $\Tilde{J}$ could be independently controlled if $t_\parallel^\alpha$ from each layer $\alpha$ of the physical bilayer are controlled individually. Note that this implies $\gamma\neq 1$, and hence a slight deviation from the original $t$-$J$ model. $(ii)$ As we will show below, the fact that doped holes in the $t$-$J$ monolayer Eq. \eqref{eq:Bilayer_to_tJ} correspond to particles in the physical bilayer model allows to measure the momentum-resolved density of holes.

\subsection{Rung singlets in the strong $J_\perp\gg t_\parallel$ limit: numerical results}

The mapping introduced above relies on the formation of singlets on each rung that represent doped holes in the effective $t$-$J$ model. This mapping is exact in the $ J_\perp/t_\parallel \to \infty$ limit. In order to estimate the effect of finite $J_\perp/t_\parallel < \infty$  as realized in experiments, we consider the case of layer-independent parameters $J_\parallel$, $t_\parallel$ and calculate the singlet expectation value for experimentally realized regimes $J_\parallel/t_\parallel=0.07 $ and $J_\perp/t_\parallel = 0.5,\dots,3.0$ \cite{Hirthe2022} using DMRG. Specifically, we consider a ladder geometry of length $L=40$ upon hole doping the lower leg $\delta=0,\dots,0.95$. Note that since the effective $t$-$J$ model \eqref{eq:Bilayer_to_tJ} is formulated w.r.t. to a vacuum of $L$ rung singlets of the actual mixD bilayer, this corresponds to hole doping the $t$-$J$ with $1-\delta$. Fig.~\ref{fig:measurement_tJ} \textbf{1)} shows the simulated system and the numerical results for the rung singlet probability averaged over all rungs and normalized by the number of particles $N_p$ present in the system, 
\begin{align}
    \mathcal{A}_S=\frac{1}{N_p}\sum_i^L \langle - \hat{\mathbf{S}}_{i0} \cdot \hat{\mathbf{S}}_{i1}+\frac{1}{4}\hat{n}_{i0}\hat{n}_{i1}\rangle.
    \label{eq:As}
\end{align}
For all $\delta$ we find  $\mathcal{A}_S>0.8$ for sufficiently large Kondo coupling $J_\perp>1.5t_\parallel$ and $\mathcal{A}_S>0.9$ for $J_\perp>2.5t_\parallel$,  i.e. singlets dominate along the rungs as required for the mixD bilayer to single-layer $t$-$J$ mapping.

\subsection{Measurement protocol}
The mixD bilayer can be realized experimentally in the large $J_\perp$ regime that is needed for the mapping to the single-layer $t$-$J$ model \eqref{eq:Bilayer_to_tJ}, e.g. with the same preparation scheme as in Ref. \cite{Hirthe2022}, where a mixed dimensional ladder of $J_\perp/J= 21(5)\gg 1$ was realized. 
Using the mapping \eqref{eq:Bilayer_to_tJ}, momentum-resolved hole densities $\braket{\hat{n}_h(\mathbf{k})}$ can be accessed through the following measurement protocol, summarized in  Fig. \ref{fig:measurement_tJ}:
\begin{enumerate}[{(1)}]
    \item After state preparation, the dynamics is frozen by ramping up a deep lattice to start the measurement procedure. 
    \item Similar to Sec. \ref{sec:measurement}, the rung singlets are brought into resonance with doublons in the lower layer for $\Delta=U$, see Fig. \ref{fig:transitions}. Using a $\pi$ tunneling pulse, singlets are transformed into doublons in the lower layer. Then, the upper layer is spatially removed from the lower $\alpha=2$ layer.
    \item  The remaining $\alpha=2$ layer consists only of doublons or empty sites. Upon removing one spin species from this layer and ramping down the intralayer lattice depth, the remaining indistinguishable particles move freely, allowing for a time-of-flight (TOF) measurement similar to Ref. \cite{Bohrdt2018_angle} in order to determine their momentum resolved density $\n_h(\vec{k})$.
\end{enumerate}
In the last step, a bandmapping \cite{Greiner2001} is applied which maps the quasi momentum states in the presence of the in-plane optical lattice potential of the form $V(x,y)=V_{xy}\left( \mathrm{cos}^2(\pi x/a_x)+\mathrm{cos}^2(\pi y/a_y)\right)$ (with lattice constants $a_\mu$) to momentum states by ramping down $V_{xy}$ \cite{Greiner2001,Bohrdt2018_angle}. The position $\vec{r}_\mathrm{TOF}$ of each atom after a free expansion of the system for a duration $t_\mathrm{TOF}$ is determined using a quantum gas microscope and mapped to its momentum $\vec{k}$ with $k_\mu = \pi m \lambda^2_\mu \vec{r}^\mu_\mathrm{TOF} /(2ht_\mathrm{TOF} )$ \cite{Bohrdt2018_angle}. Note that in order to achieve a sufficiently long time-of-flight, the initial in-plane system size $L_\mu$ has to be small compared to the total system size $L^\mathrm{TOF}_\mu$. In particular, $L^\mathrm{TOF}_\mu/L_\mu$ limits the momentum resolution in direction $\mu=x,y$. 

As a slight modification of the above protocol, we note that it is also possible to prepare a doublon-doped FH system in the (initially) energetically lower layer, while the other one is empty. Doublons are then transferred to the other layer via two consecutive $\pi$-pulses (where in the second step the sign of the energy offset is switched). In analogy to the above scheme, one spin-species is removed to ensure indistinguishability and TOF measurements can be performed. This way, the implementation of strong interlayer couplings $J_{\perp}$ can be avoided; however, virtual doublon-hole fluctuations may have notable effects for moderately strong repulsions. 

\section{Discussion}
Motivated by recent studies of high-temperature superconductivity in bilayer nickelates~\cite{Sun2023, zhang2023_zeroR, hou2023emergence}, we have proposed a scheme to prepare and measure a state with (quasi) long-range pair coherence in ultracold atom simulators. With estimated critical temperatures on the order of $J_{\perp}/2$ in the mixD bilayer $t$-$J$ model, our scheme provides a directly realizable protocol for the observation of superconducting correlations in a system with strongly repulsively interacting fermions with current state-of-the-art experimental platforms. 
Additionally, simulating minimal models of bilayer nickelates may allow for an experimental observation of BEC-BCS-type crossovers as a function of $t_{\parallel}/J_{\perp}$~\cite{schlömer2023superconductivity, yang2023strong, lu2023superconductivity}. Furthermore, differences between single- and multi-band models can be studied by engineering tailored bilayer and ladder systems that mimic multiple orbitals. This opens the door to directly simulate strongly correlated materials with ultracold atoms in optical lattices, and may be a major step towards designing new materials with high critical temperatures.

Moreover, local addressability of spin-flip and tunneling gates allows to measure both coherent pairing correlations and momentum-resolved dopant densities in 2D FH models. While the former utilizes a mapping of the repulsive FH to its closely related attractive cousin,   the latter is achieved by projecting mobile holes to interlayer singlets, whose momentum can be accessed through time-of-flight measurements in the auxiliary layer. These capabilities directly bridge solid-state and cold atom experiments, and may help to microscopically study the enigmatic (normal and superconducting) phases appearing in copper-oxide compounds. Our work also paves the way for a direct measure of $d$-wave pairing fluctuations in the plain-vanilla Hubbard model using ultracold atoms in optical lattices.\\

\textbf{Acknowledgements.} We thank Eugene Demler, Lukas Homeier, Christian Kokail and Ulrich Schollwöck for insightful discussions. This research was funded by the Deutsche Forschungsgemeinschaft (DFG, German Research Foundation) under Germany’s Excellence Strategy—EXC-2111—390814868 and by the European Research Council (ERC) under the European Union’s Horizon 2020 research and innovation programme (grant agreement number 948141 — ERC Starting Grant SimUcQuam). HL acknowledges support by the International Max Planck Research School for Quantum Science and Technology (IMPRS-QST).

\onecolumngrid
\widetext

\appendix



\section{Pairing correlations in mixed-dimensional bilayers}
\subsection{Particle-hole transformation}
\label{sec:A1}
We here review the motivation for the particle-hole transformation Eq.~\eqref{eq:trafo} (see also e.g.~\cite{Altland_Simons_2010}). Consider for this the charge conjugation transformation $C$, which maps particle creation to annihilation operators and vice versa,
\begin{equation}
\begin{gathered}
    C \hat{c}_{\mathbf{i}, \sigma} C^{-1} =  \hat{c}_{\mathbf{i}, \sigma}^{\dagger} \\
    C \hat{c}_{\mathbf{i}, \sigma}^{\dagger} C^{-1} = \hat{c}_{\mathbf{i}, \sigma}.
\end{gathered}
\label{eq:trafo_cg}
\end{equation}
Let us evaluate how the single particle states $\{\ket{0}, \ket{\uparrow}, \ket{\downarrow}\}$ behave under the transformation $C$. First, consider the action on the vacuum state, which is defined by $\hat{c}_{\mathbf{i}, \sigma}\ket{0} = 0$ for all $\mathbf{i}, \sigma$. Applying the transformation yields 
\begin{equation}
0 = C \hat{c}_{\mathbf{i}, \sigma} \ket{0} = C \hat{c}_{\mathbf{i}, \sigma} C^{-1} C \ket{0} = \hat{c}^{\dagger}_{\mathbf{i}, \sigma} C 
\ket{0},
\end{equation}
such that the transformed vacuum state is the fully occupied state, $C \ket{0} = \prod_{\mathbf{i}, \sigma} c^{\dagger}_{\mathbf{i}, \sigma} \ket{0}$. Furthermore, we get for each $\mathbf{i}$ (we omit the lattice site index $\mathbf{i}$ for simplicity)
\begin{equation}
    \begin{aligned}
        C \ket{\uparrow} = C \hat{c}^{\dagger}_{\uparrow} \ket{0} = C \hat{c}^{\dagger}_{\uparrow} C^{-1} C \ket{0} = \hat{c}^{\vphantom\dagger}_{\uparrow} \hat{c}^{\dagger}_{\uparrow} \hat{c}^{\dagger}_{\downarrow} \ket{0} = \hat{c}^{\dagger}_{\downarrow} \ket{0} = \ket{\downarrow} 
    \end{aligned}
\end{equation}
and 
\begin{equation}
    \begin{aligned}
        C \ket{\downarrow} = C \hat{c}^{\dagger}_{\downarrow} \ket{0} = C \hat{c}^{\dagger}_{\downarrow} C^{-1} C \ket{0} = \hat{c}^{\vphantom\dagger}_{\downarrow} \hat{c}^{\dagger}_{\uparrow} \hat{c}^{\dagger}_{\downarrow} \ket{0} = -\hat{c}^{\dagger}_{\uparrow} \ket{0} = -\ket{\uparrow}.    
    \end{aligned}
\end{equation}
The spin flips of the single particle states under the transformation as seen above are intuitive when considering a state in the subspace of single and double occupancies, $\ket{\uparrow, \downarrow, \uparrow\downarrow, \uparrow}$. Applying the hopping term $\hat{c}_{2, \uparrow}^{\dagger} \hat{c}_{3, \uparrow}^{\vphantom\dagger} \ket{\uparrow, \downarrow, \uparrow\downarrow, \uparrow} = - \ket{\uparrow, \uparrow \downarrow, \downarrow, \uparrow}$, we see that the hopping of the $\uparrow$-spin maps to a hopping of a $\downarrow$-spin in the subspace of empty and singly occupied sites, $\hat{c}_{3, \downarrow}^{\dagger} \hat{c}_{2, \downarrow}^{\vphantom\dagger} \ket{\uparrow, \downarrow, 0, \uparrow} = \ket{\uparrow, 0, \downarrow, \uparrow}$. Note that there is an additional sign change of the hopping term after the transformation, as $C\hat{c}^{\dagger}_{\mathbf{i}, \sigma} \hat{c}^{\vphantom\dagger}_{\mathbf{j}, \sigma} C^{-1} = \hat{c}^{\vphantom\dagger}_{\mathbf{i}, \sigma} \hat{c}^{\dagger}_{\mathbf{j}, \sigma} = - \hat{c}^{\dagger}_{\mathbf{j}, \sigma} \hat{c}^{\vphantom\dagger}_{\mathbf{i}, \sigma}$ (for $\mathbf{i} \neq \mathbf{j}$).  

To account for the appearing spin and phase flips, we redefine the charge conjugation operation $C$ and make it site- and spin-dependent (i.e., we add a unitary transformation to the charge conjugation Eq.~\eqref{eq:trafo_cg}, $\mathcal{U}C \mathcal{U}^{\dagger} = \hat{\mathcal{C}}$, representing another possible particle-hole transformation),
\begin{equation}
\begin{gathered}
    \hat{\mathcal{C}} \hat{c}_{\mathbf{i}, \sigma} \hat{\mathcal{C}}^{-1} = \eta_{\mathbf{i}} \text{sgn}(\bar{\sigma}) \hat{c}_{\mathbf{i}, \bar{\sigma}}^{\dagger} \\
    \hat{\mathcal{C}} \hat{c}_{\mathbf{i}, \sigma}^{\dagger} \hat{\mathcal{C}}^{-1} = \eta_{\mathbf{i}} \text{sgn}(\bar{\sigma}) \hat{c}_{\mathbf{i}, \bar{\sigma}}.
\end{gathered}
\end{equation}
Here, the sign factor $\eta_{\mathbf{i}} = e^{i \boldsymbol{\pi} \cdot \mathbf{i}}$ with $\boldsymbol{\pi} = [\pi, \pi]$ is positive (negative) on sublattice A (B) on the square lattice; note that $\hat{\mathcal{C}}$ also switches spins $\sigma \leftrightarrow \bar{\sigma}$. 

Applying the transformation to single particle states on a given site ($\hat{\mathcal{C}} \ket{0} = \prod_{\mathbf{i}, \sigma} \hat{c}^{\dagger}_{\mathbf{i}, \sigma} \ket{0}$ still holds),
\begin{equation}
    \begin{aligned}
        &\hat{\mathcal{C}} \ket{\uparrow} = \hat{\mathcal{C}} \hat{c}^{\dagger}_{\uparrow} \hat{\mathcal{C}}^{-1} \hat{\mathcal{C}} \ket{0} = - \hat{c}^{\vphantom\dagger}_{\downarrow} \hat{c}^{\dagger}_{\uparrow} \hat{c}^{\dagger}_{\downarrow} \ket{0} = \ket{\uparrow} \\
        &\hat{\mathcal{C}} \ket{\downarrow} = \hat{\mathcal{C}} \hat{c}^{\dagger}_{\downarrow} \hat{\mathcal{C}}^{-1} \hat{\mathcal{C}} \ket{0} = \phantom{-} \hat{c}^{\vphantom\dagger}_{\uparrow} \hat{c}^{\dagger}_{\uparrow} \hat{c}^{\dagger}_{\downarrow} \ket{0} = \ket{\downarrow},   
    \end{aligned}
\end{equation}
such that the singly occupied states map onto themselves, while doublons map to holes and vice versa.
Transforming the relevant operators in the Hamiltonian Eq.~\eqref{eq:tJp} for nearest neighbor pairs $\braket{\mathbf{i}, \mathbf{j}}$ yields
\begin{equation}
    \begin{gathered}
        \hat{\mathcal{C}} \hat{n}_{\mathbf{i}, \sigma} \hat{\mathcal{C}}^{-1} = \hat{c}_{\mathbf{i}, \bar{\sigma}} \hat{c}^{\dagger}_{\mathbf{i}, \bar{\sigma}} =  1 - \hat{n}_{\mathbf{i}, \bar{\sigma}} \\
        \hat{\mathcal{C}} \hat{n}_{\mathbf{i}} \hat{\mathcal{C}}^{-1} = \hat{\tilde{n}}_{\mathbf{i}} \\
        \hat{\mathcal{C}} \hat{\mathcal{P}} \hat{\mathcal{C}}^{-1} = \hat{\tilde{\mathcal{P}}} \\
        \hat{\mathcal{C}} \hat{c}_{\mathbf{i}, \sigma}^{\dagger} \hat{c}_{\mathbf{j}, \sigma}^{\vphantom\dagger} \hat{\mathcal{C}}^{-1} = - \hat{c}_{\mathbf{i}, \bar{\sigma}}^{\vphantom\dagger} \hat{c}_{\mathbf{j}, \bar{\sigma}}^{\dagger} = \hat{c}_{\mathbf{j}, \bar{\sigma}}^{\dagger} \hat{c}_{\mathbf{i}, \bar{\sigma}}^{\vphantom\dagger} \\
        \hat{\mathcal{C}} \hat{S}^z_{\mathbf{i}} \hat{\mathcal{C}}^{-1} = \frac{1}{2} \hat{\mathcal{C}} (\hat{n}_{\mathbf{i}, \uparrow} - \hat{n}_{\mathbf{i}, \downarrow}) \hat{\mathcal{C}}^{-1} = \hat{S}^z_{\mathbf{i}} \\
        \hat{\mathcal{C}} \hat{S}^x_{\mathbf{i}} \hat{\mathcal{C}}^{-1} =\frac{1}{2} \hat{\mathcal{C}} (\hat{c}^{\dagger}_{\mathbf{i}, \uparrow} \hat{c}^{\vphantom\dagger}_{\mathbf{i}, \downarrow} + \hat{c}^{\dagger}_{\mathbf{i}, \downarrow} \hat{c}^{\vphantom\dagger}_{\mathbf{i}, \uparrow}) \hat{\mathcal{C}}^{-1} = \hat{S}^x_{\mathbf{i}} \\     
        \hat{\mathcal{C}} \hat{S}^y_{\mathbf{i}} \hat{\mathcal{C}}^{-1} = \frac{1}{2} i\hat{\mathcal{C}} ( \hat{c}^{\dagger}_{\mathbf{i}, \downarrow} \hat{c}^{\vphantom\dagger}_{\mathbf{i}, \uparrow}- \hat{c}^{\dagger}_{\mathbf{i}, \uparrow} \hat{c}^{\vphantom\dagger}_{\mathbf{i}, \downarrow} ) \hat{\mathcal{C}}^{-1} = \hat{S}^y_{\mathbf{i}} 
    \end{gathered}
    \label{eq:trafo_coll}
\end{equation}
Furthermore, the total particle number transforms as $\hat{\mathcal{C}} \hat{N} \hat{\mathcal{C}}^{-1} = \sum_{\mathbf{i}} \Big( 2 - \hat{n}_{\mathbf{i}, \uparrow} - \hat{n}_{\mathbf{i}, \downarrow}\Big) = L - d$, such that $\sum_{\mathbf{i}} \hat{n}_{\mathbf{i}, \uparrow} + \hat{n}_{\mathbf{i}, \downarrow} = L + d$, i.e., the transformed system has a total of $L+d$ particles ($d$ doublons). 

\subsection{Charge gaps \label{appendix:Gaps}}
\begin{figure*}
\centering
\includegraphics[width=0.95\textwidth]{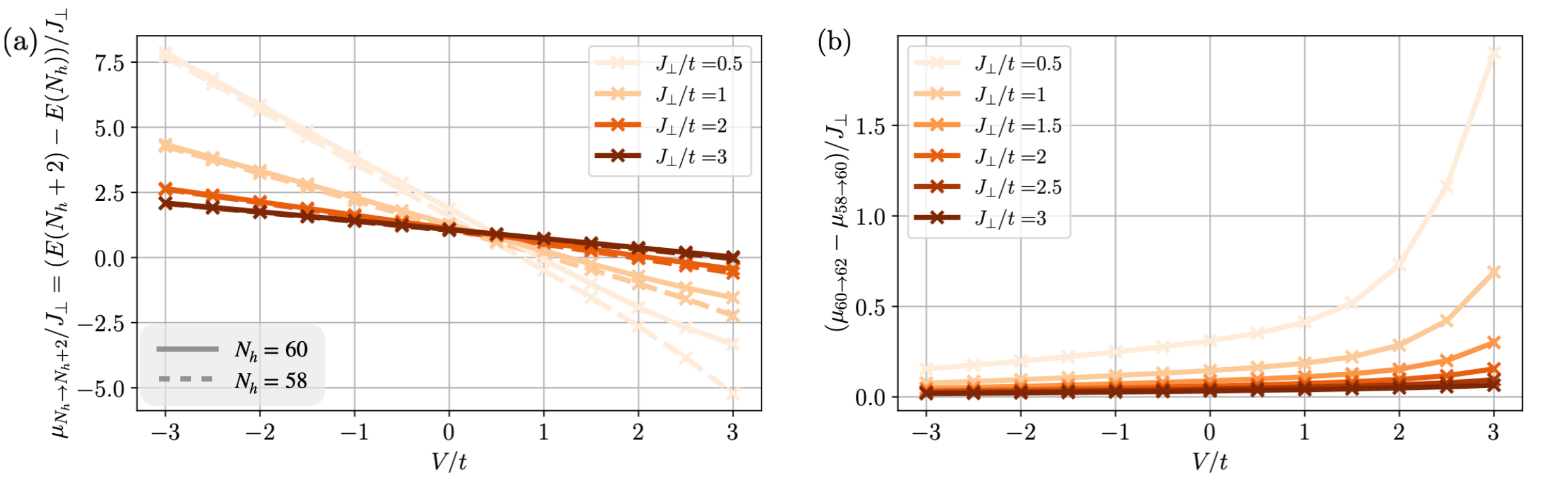}
\caption{\textbf{DMRG simulations of charge gaps.} (a) $\mu_{N_h+2\to Nh}=E(N_h+2)-E(N_h)$ in the ground state of a mixD ladder of length $L_x=60$ at hole doping $N_h=60$ ($\delta=0.50$, solid lines) and $N_h=58$ ($\delta=0.48$, dashed lines), for $J_\parallel/t_{\parallel}=0.5$  and for various Kondo couplings $J_\perp/t$ and interaction strengths $V/t_{\parallel}$. (b) Differences $\mu_{N_h+2=62\to Nh=60}-\mu_{N_h+2=60\to Nh=58}$. In both plots the opening of a charge gap for large $V$ is clearly visible.}
\label{fig:Gaps}
\end{figure*}
In Sec. \ref{sec:mixDbilayer}, Fig. \ref{fig:DeltaiDeltaj} we observe long-range pairing correlations in the mixD bilayer, with the onset of an exponential decay for large $V$ and small $J_\perp$. This is accompanied with the opening of a charge gap $\mu_{N_h+2\to Nh}$ at the respective values of $V$ and small $J_\perp$, as shown for the same ladder systems of length $L_x=60$ in Fig. \ref{fig:Gaps}: In Fig. \ref{fig:Gaps}~(a), we show $\mu_{N_h+2\to Nh}=E(N_h+2)-E(N_h)$ in the ground state of a mixD ladder of length $L_x=60$ at hole doping $N_h=60$ ($\delta=0.50$) and $N_h=58$ ($\delta=0.48$). At commensurate $\delta=0.50$, $\mu_{N_h+2\to Nh}$ is significantly decreased w.r.t $\delta=0.48$ for large $V/t$. The strongest decrease is observed for small $J_\perp/t=0.5$ as soon as $V/t$ becomes large $V/t>1$, and the smallest decrease is found for large $J_\perp/t=3$. This is in agreement with the onset of an exponential decay of the pairing correlations only for small $J_\perp/t$ and large $V/t$. The decrease is explicitly shown in Fig. \ref{fig:Gaps}~(b).

\subsection{Distinguishing singlet from triplet states}
\label{appendix:sto_adia}
In order to measure coherent pairing correlations as described in Sec.~\ref{sec:measurement}, singlet states need to be distinguished from triplet states after ramping up the lattice depths and freezing out in-plane motion. In order to do so, we propose to adiabatically ramp up a magnetic field gradient, which has the following effect: Consider the Hamiltonian 
\begin{equation}\label{eq:H_STO}\hat{\mathcal{H}}(B_z) = \hat{\mathcal{H}}_{\text{FH}}(t,U) + B_z \hat{S}^z_2,\end{equation} where $\hat{\mathcal{H}}_{\text{FH}}(t,U)$ is the FH model with hopping (on-site repulsion) $t$ ($U$), and $B_z$ denotes the magnetic field gradient strength. If $B_z$ is ramped up adiabatically, $\ket{t_1} = \ket{\downarrow \downarrow}$ and $\ket{t_3} = \ket{\uparrow \uparrow}$ stay eigenstates of the Hamiltonian throughout the transformation. On the other hand, the singlet state will adiabatically transform to $\ket{s} \rightarrow \ket{\uparrow \downarrow}$, and the triplet state $\ket{t_2}$ transforms to the remaining product state $\ket{\downarrow \uparrow}$. This is illustrated in the left panel of Fig.~\ref{fig:STO_adiabatic}, where we show the instantaneous eigenenergies of $\hat{\mathcal{H}}(B_z)$ as a function of $B_z/t$ for $U/t = 12$. Spin-resolved snapshots after rapidly turning off the field then allow to distinguish between states that have originally been in a spin-singlet and spin-triplet state. We note that (after freezing in-plane dynamics) the Hamiltonian conserves $S_z^{\text{tot}}$ in each double well, such that diabatic transitions to other states in the triplet manifold, e.g. $\ket{\uparrow \downarrow} \rightarrow \ket{\downarrow \downarrow}$, are suppressed.

We note that the adiabatic process is in contrast to singlet-triplet oscillations, where the state is quenched under the magnetic field gradient Hamiltonian (i.e., the process is non-adiabatic). In this case, the relative phase between the states $\ket{\uparrow \downarrow}$ and $\ket{\downarrow \uparrow}$ is dynamically changed, $\ket{\uparrow \downarrow} + e^{i\varphi(t)} \ket{\downarrow \uparrow}$.

\section{2D Fermi-Hubbard model}
\subsection{$\hat{U}_{B_z}$ gate}
\label{appendix:UBz}
\begin{figure*}
\centering
\includegraphics[width=0.65\textwidth]{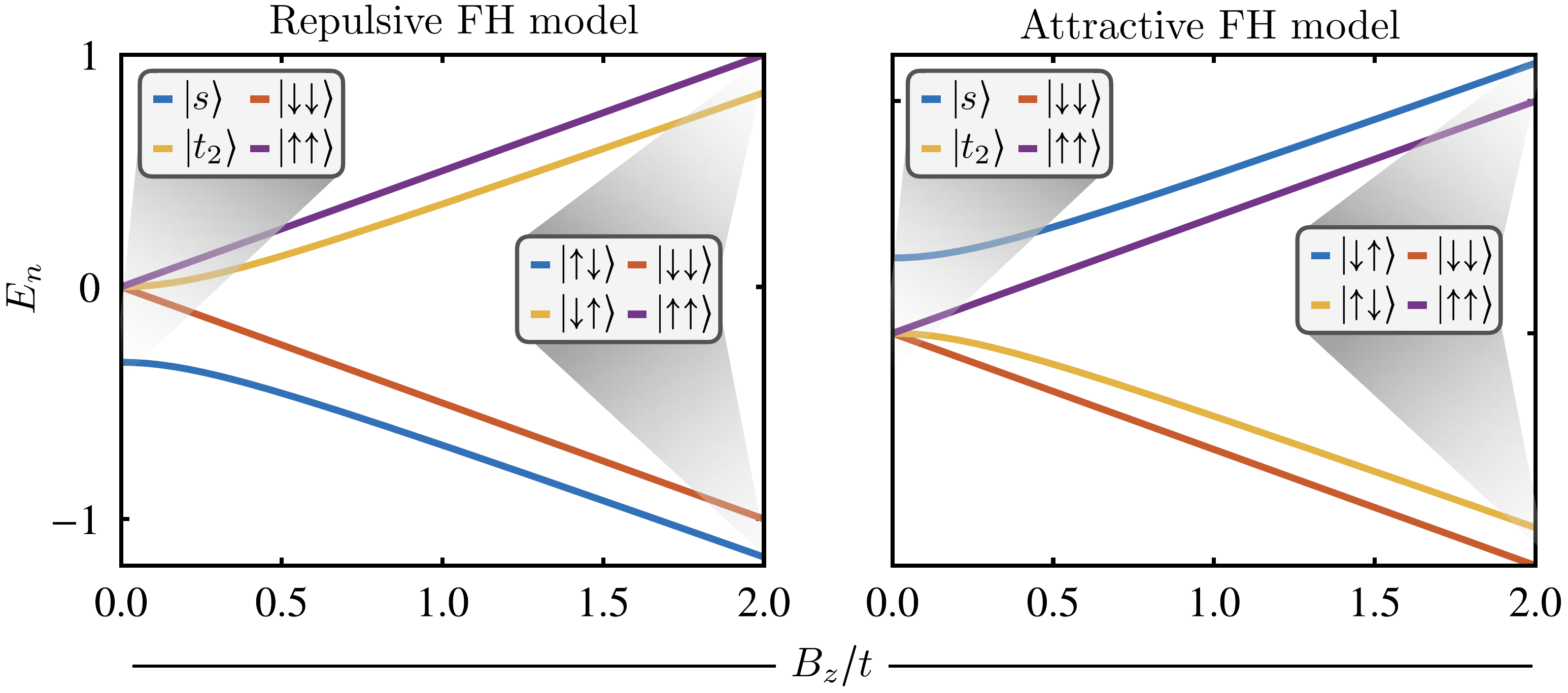}
\caption{\textbf{Adiabatic magnetic field gradient ramp.} Evolution of instantaneous eigenstate energies of Hamiltonian Eq.~\eqref{eq:H_STO} as a function of $B_z/t$, with $U/t = 12$ ($U/t=-12$) on the left-hand (right-hand) side. For $B_z/t = 0$, triplet states are degenerate and energetically offset from the singlet state. While $\ket{\uparrow \uparrow}$ and $\ket{\downarrow \downarrow}$ stay invariant under the adiabatic transformation, the singlet state transforms to $\ket{\uparrow \downarrow}$ ($\ket{\downarrow \uparrow}$), and the triplet state to $\ket{\downarrow \uparrow}$ ($\ket{\uparrow \downarrow}$) in the repulsive (attractive) case. This allows to $(i)$ distinguish singlet from triplet states in spin-resolved snapshots, as required in measurement scheme Sec.~\ref{sec:measurement}, and $(ii)$ apply the transformation $\ket{\uparrow\downarrow}\rightarrow\ket{t_2}$ through a reversed adiabatic protocol as proposed in measurement scheme Sec.~\ref{sec:U}.}
\label{fig:STO_adiabatic}
\end{figure*}

During the measurement gate sequence, the state $\ket{\uparrow \downarrow}$ is transformed into the triplet $\hat{U}_{B_z} \ket{\uparrow \downarrow} = \ket{t_2} = \nicefrac{1}{\sqrt{2}} (\ket{\uparrow \downarrow} + \ket{\downarrow \uparrow})$. This can be realized through a reversed adiabatic protocol as described in Appendix~\ref{appendix:sto_adia}: A magnetic field gradient is rapidly turned on, and then ramped down adiabatically to $B_z \rightarrow 0$. The instantaneous eigenenergies of the double-well Hamiltonian Eq.~\eqref{eq:H_STO} as a function of $B_z$ are shown for $U/t = -12$ in the right panel of Fig.~\ref{fig:STO_adiabatic}. During the adiabatic process, $\ket{\uparrow \downarrow} \rightarrow \ket{t_2}$, while the doublon-hole state $\ket{\uparrow\downarrow,0}$ remains unaffected as required in the measurement scheme Sec.~\ref{sec:U}. 

\subsection{Mixed states}
\label{appendix:2DHubbard}
We here show that the arguments presented in Sec.~\ref{sec:FH_pairing} to access pairing correlations in the 2D FH model hold also for a general correlated state. In particular, expectation values are given by $\braket{\hat{\Delta}^{\dagger}_{\mathbf{r
}_1} \hat{\Delta}_{\mathbf{r
}_2}} = \text{tr}\left(\hat{\rho}_{12}^{sh} \hat{\Delta}^{\dagger}_{\mathbf{r
}_1} \hat{\Delta}_{\mathbf{r
}_2}\right)$, where $\hat{\rho}_{12}^{sh}$ is the reduced density matrix in the subspace of hole-pairs and singlets on bonds $\mathbf{r}_1, \mathbf{r}_2$ (as $\hat{\Delta}^{(\dagger)}_{\mathbf{r
}_i}$ only acts on this subspace). The most general form of $\hat{\rho}_{12}^{sh}$ in the basis $\{\ket{00}_1\ket{00}_2, \ket{s}_1\ket{00}_2, \ket{00}_1\ket{s}_2, \ket{s}_1\ket{s}_2 \}$ reads (note that fermion number conservation implies $a_1=a_2=b_1=b_2=c_2=0$),
\begin{equation}
    \hat{\rho}_{12}^{sh} = \begin{bmatrix} p_1& a_2^* & a_1^* & c_2^*\\ a_2 & p_2 & c_1^* & b_1^* \\ a_1 & c_1 & p_3 & b_2^* \\ c_2 & b_1 & b_2 & p_4
    \end{bmatrix},
\end{equation}
with $\sum_{j=1}^4 p_j = 1$. Thus, pairing correlations are given by the anti-diagonal elements, 
\begin{equation}
    \text{tr}\left[\hat{\rho}_{12}^{sh} \big(\hat{\Delta}^\dagger_{\mathbf{r}_1} + \hat{\Delta}_{\mathbf{r}_1}\big)\big(\hat{\Delta}^\dagger_{\mathbf{r}_2} + \hat{\Delta}_{\mathbf{r}_2}\big)\right] = c_1 + c_1^* + c_2 + c_2^*.
\end{equation}
As we show now, these elements can be accessed by taking snapshots in the rotated basis (i.e. after applying the attractive-to-repulsive $U$ mapping and the unitary $\hat{U}^{\varphi}$). Specifically, in analogy to the strategy for a product state as outlined in the main text, we calculate the probability of measuring the transformed state as $\ket{\uparrow \downarrow}_1 \ket{\uparrow \downarrow}_2$,
\begin{equation}
    P^{\uparrow\downarrow, \uparrow\downarrow}_{\mathbf{r}_1, \mathbf{r}_2} = \text{tr}\left[ \hat{\rho}_{12}^{sh} \hat{U}^{\varphi} \ket{\uparrow\downarrow}_1 \ket{\uparrow\downarrow}_2 \bra{\uparrow\downarrow}_2 \bra{\uparrow\downarrow}_1 \hat{U}^{\varphi \dagger} \right].
\end{equation}
For $\varphi = 0$, the transformed state $\hat{U}^{\varphi=0}\ket{\uparrow\downarrow}_1\ket{\uparrow\downarrow}_2 \propto \hat{U}^{\varphi=0} ( \ket{t_2}_1\ket{t_2}_2 + \ket{t_2}_1\ket{s}_2 + \ket{s}_1\ket{t_2}_2 +\ket{s}_1\ket{s}_2)$ corresponds to the equal superposition $\ket{00}_1\ket{00}_2 + \ket{00}_1\ket{s}_2 + \ket{s}_1\ket{00}_2 +\ket{s}_1\ket{s}_2$ in the repulsive model. Extending this to the other possible states in the rotated basis, we find 
\begin{equation}
\begin{aligned}
\hat{U}^{\varphi=0}\ket{\uparrow\downarrow}_1\ket{\uparrow\downarrow}_2 \hat{\sim} \ket{00}_1\ket{00}_2 + \ket{00}_1\ket{s}_2 + \ket{s}_1\ket{00}_2 +\ket{s}_1\ket{s}_2, \\
\hat{U}^{\varphi=0}\ket{\uparrow\downarrow}_1\ket{\downarrow\uparrow}_2 \hat{\sim} \ket{00}_1\ket{00}_2 + \ket{00}_1\ket{s}_2 - \ket{s}_1\ket{00}_2 -\ket{s}_1\ket{s}_2, \\
\hat{U}^{\varphi=0}\ket{\downarrow\uparrow}_1\ket{\uparrow\downarrow}_2 \hat{\sim} \ket{00}_1\ket{00}_2 - \ket{00}_1\ket{s}_2 + \ket{s}_1\ket{00}_2 -\ket{s}_1\ket{s}_2, \\
\hat{U}^{\varphi=0}\ket{\downarrow\uparrow}_1\ket{\downarrow\uparrow}_2 \hat{\sim} \ket{00}_1\ket{00}_2 - \ket{00}_1\ket{s}_2 - \ket{s}_1\ket{00}_2 +\ket{s}_1\ket{s}_2.
\end{aligned}
\end{equation}
From this, it is straight-forward to show that 
\begin{equation}
P_{\mathbf{r}_1, \mathbf{r}_2}^{\uparrow\downarrow,\uparrow\downarrow} + P_{\mathbf{r}_1, \mathbf{r}_2}^{\downarrow\uparrow, \downarrow\uparrow} - P_{\mathbf{r}_1, \mathbf{r}_2}^{\downarrow\uparrow, \uparrow\downarrow}-P_{\mathbf{r}_1, \mathbf{r}_2}^{\uparrow\downarrow,
\downarrow\uparrow} \sim c_1 + c_1^* + c_2 + c_2^*,
\end{equation}
which extends the results presented in the main text to arbitrary correlated many body states. 

\section{MixD bilayer to single-layer $t$-$J$ model mapping \label{appendix:2DtJ}}
Here, we consider the mapping of the mixD bilayer, with microscopic fermions with spin-index $\sigma$ and layer-index $\alpha$ represented by $\C_{\mathbf{i},\sigma,\alpha}$, to a single-layer $t$-$J$ model in the $U,J_\perp \gg t_\parallel$ limit and derive the effective Hamiltonian \eqref{eq:Bilayer_to_tJ}. In the large $U \gg t_\parallel$ limit, doubly occupied sites are projected out and we get the anti-commutation relations, see e.g. Ref. \cite{Batista_2004}, 
\begin{align}
    \{\Cd_{\mathbf{i},\sigma,\alpha},\Cd_{\mathbf{j},\sigma^\prime,\alpha^\prime}\} = \{\C_{\mathbf{i},\sigma,\alpha},\C_{\mathbf{j},\sigma^\prime,\alpha^\prime}\}=0,
\end{align}
and 
\begin{align}
    \{\C_{\mathbf{i},\sigma,\alpha},\Cd_{\mathbf{j},\sigma^\prime,\alpha^\prime}\} = \delta_{\mathbf{i}\mathbf{j}}\delta_{\alpha\alpha^\prime}\left(\delta_{\sigma \sigma^\prime}(1-\n_{\mathbf{i}\alpha})+\Cd_{\mathbf{i\sigma^\prime,\alpha^\prime}}\C_{\mathbf{i},\sigma,\alpha}\right),
\end{align}
The latter implies that no opposite spin can be created on top of another spin since  $\C_{\mathbf{i},\uparrow,\alpha},\Cd_{\mathbf{i},\uparrow,\alpha}\ket{\downarrow}=(1-1+\C_{\mathbf{i},\uparrow,\alpha}\Cd_{\mathbf{i},\uparrow,\alpha}-\C_{\mathbf{i},\uparrow,\alpha}\Cd_{\mathbf{i},\uparrow,\alpha})\ket{\downarrow}=0$ and $\C_{\mathbf{i},\downarrow,\alpha},\Cd_{\mathbf{i},\uparrow,\alpha}\ket{\downarrow}=0$, reflecting the single-occupancy constraint. Furthermore, $\C_{\mathbf{i},\downarrow,\alpha},\Cd_{\mathbf{i},\uparrow,\alpha}\ket{0}=0$ and $\C_{\mathbf{i},\uparrow,\alpha},\Cd_{\mathbf{i},\uparrow,\alpha}\ket{0}=\ket{0}$.

\subsection{Derivation of fermionic commutations}
In the following we derive the fermionic anti-commutation relation for the operators $\hat{\Tilde{c}}_{\mathbf{i},\sigma}$ of the effective model,
\begin{equation}
    \begin{aligned}
    \{\hat{\Tilde{c}}_{\mathbf{i}\sigma} ,\hat{\Tilde{c}}_{\mathbf{j}\sigma^\prime}^\dagger\} = \delta_{\mathbf{i}\mathbf{j}}\left[\delta_{\sigma \sigma^\prime}\Bd_\mathbf{i}\B_\mathbf{i} + \Cd_{\mathbf{i},\sigma^\prime,1}\C_{\mathbf{i},\sigma,1}(1-\n_{\mathbf{i},2})\right].
    \label{eq:newc_anticommutation}
    \end{aligned}
\end{equation}

The commutation relations for the bosonic and fermionic parts of $\hat{\Tilde{c}}^\dagger_{\mathbf{i},\sigma}=\Cd_{\mathbf{i},\sigma,1} \B_{\mathbf{i}}$  are as follows:
\begin{align}
    \left[ \B_\mathbf{i}, \Bd_\mathbf{j} \right] = \frac{1}{2} \left[\C_{\mathbf{i},\downarrow,2}\C_{\mathbf{i},\uparrow,1}-\C_{\mathbf{i},\uparrow,2}\C_{\mathbf{i},\downarrow,1}, \Cd_{\mathbf{j},\uparrow,1} \Cd_{\mathbf{j},\downarrow,2}-\Cd_{\mathbf{j},\downarrow,1} \Cd_{\mathbf{j},\uparrow,2}\right] = \delta_{\mathbf{i}\mathbf{j}}(1-\n_{\mathbf{i},1}-\n_{\mathbf{i},2}+\n_{\mathbf{i},2}\n_{\mathbf{i},1}-\Bd_\mathbf{i}\B_\mathbf{i}).
    \label{eq:bb}
\end{align}
Eq. \eqref{eq:bb} is the commutation relation for hard core bosons. In particular 
$$  \langle \singlet \vert \B_\mathbf{i} \Bd_\mathbf{i}\vert\singlet \rangle = \langle \singlet \vert  1-\n_{\mathbf{i},1}-\n_{\mathbf{i},2}+\n_{\mathbf{i},2}\n_{\mathbf{i},1} \vert\singlet \rangle=0,$$
i.e. if the rung is already occupied by a singlet it is not possible to create another singlet on the same rung. Furthermore, Eq. \eqref{eq:bb} implies $ \langle \holehole \vert  \B_\mathbf{i} \Bd_\mathbf{i}\vert\holehole \rangle =1$ and $ \langle \holesigma \vert  \B_\mathbf{i} \Bd_\mathbf{i}\vert\holesigma \rangle =0$. Eq. \eqref{eq:bb} is derived by calculating
\begin{align*}
 \left[\C_{\mathbf{i},\downarrow,2}\C_{\mathbf{i},\uparrow,1}, \Cd_{\mathbf{j},\uparrow,1} \Cd_{\mathbf{j},\downarrow,2}\right] &= \C_{\mathbf{i},\downarrow,2}\left[\C_{\mathbf{i},\uparrow,1}, \Cd_{\mathbf{j},\uparrow,1} \Cd_{\mathbf{j},\downarrow,2} \right] + \left[\C_{\mathbf{i},\downarrow,2}, \Cd_{\mathbf{j},\uparrow,1} \Cd_{\mathbf{j},\downarrow,2}\right]\C_{\mathbf{i},\uparrow,1} 
 \\
 &= \C_{\mathbf{i},\downarrow,2}\{\C_{\mathbf{i},\uparrow,1}, \Cd_{\mathbf{j},\uparrow,1} \}\Cd_{\mathbf{j},\downarrow,2}  - \Cd_{\mathbf{j},\uparrow,1} \{\C_{\mathbf{i},\downarrow,2}, \Cd_{\mathbf{j},\downarrow,2}\}\C_{\mathbf{i},\uparrow,1} 
 \\
 &= \delta_{\mathbf{i}\mathbf{j}}\left[\C_{\mathbf{i},\downarrow,2}(1-\n_{\mathbf{i,\downarrow,1}})\Cd_{\mathbf{j},\downarrow,2}  - \Cd_{\mathbf{j},\uparrow,1} (1-\n_{\mathbf{i},2}+\n_{\mathbf{i},\downarrow,2})\C_{\mathbf{i},\uparrow,1} \right]
  \\
 &= \delta_{\mathbf{i}\mathbf{j}}\left[(1-\n_{\mathbf{i},2})(1-\n_{\mathbf{i,\downarrow,1}})  - \Cd_{\mathbf{j},\uparrow,1} (1-\n_{\mathbf{i},2}+\n_{\mathbf{i},\downarrow,2})\C_{\mathbf{i},\uparrow,1} \right]
 \\
 &= \delta_{\mathbf{i}\mathbf{j}}\left[(1-\n_{\mathbf{i},1}-\n_{\mathbf{i},2}+\n_{\mathbf{i},2}\n_{\mathbf{i},1}-\n_{\mathbf{i},\downarrow,2}\n_{\mathbf{i},\uparrow,1})\right]
\end{align*}
and
\begin{align*}
   \left[\C_{\mathbf{i},\uparrow,2}\C_{\mathbf{i},\downarrow,1}, \Cd_{\mathbf{j},\uparrow,1} \Cd_{\mathbf{j},\downarrow,2}\right] &= \C_{\mathbf{i},\uparrow,2}\left[\C_{\mathbf{i},\downarrow,1}, \Cd_{\mathbf{j},\uparrow,1} \Cd_{\mathbf{j},\downarrow,2} \right] + \left[\C_{\mathbf{i},\uparrow,2}, \Cd_{\mathbf{j},\uparrow,1} \Cd_{\mathbf{j},\downarrow,2}\right]\C_{\mathbf{i},\downarrow,1} \\&= \C_{\mathbf{i},\uparrow,2} \{\C_{\mathbf{i},\downarrow,1}, \Cd_{\mathbf{j},\uparrow,1}\}\Cd_{\mathbf{j},\downarrow,2}-\Cd_{\mathbf{j},\uparrow,1}\{\C_{\mathbf{i},\uparrow,2}, \Cd_{\mathbf{j},\downarrow,2}\} \C_{\mathbf{i},\downarrow,1}\\
 &=\delta_{\mathbf{i}\mathbf{j}}\left( \C_{\mathbf{i},\uparrow,2} \Cd_{\mathbf{j},\uparrow,1}\C_{\mathbf{i},\downarrow,1} \Cd_{\mathbf{j},\downarrow,2}-\Cd_{\mathbf{j},\uparrow,1}\Cd_{\mathbf{j},\downarrow,2}\C_{\mathbf{i},\uparrow,2}  \C_{\mathbf{i},\downarrow,1}\right)\\
 &
 =  \delta_{\mathbf{i}\mathbf{j}}\left(\Cd_{\mathbf{j},\uparrow,1}\underbrace{\C_{\mathbf{i},\uparrow,2}\Cd_{\mathbf{j},\downarrow,2}}_{=0}\C_{\mathbf{i},\downarrow,1} -\Cd_{\mathbf{j},\uparrow,1}\Cd_{\mathbf{j},\downarrow,2}\C_{\mathbf{i},\uparrow,2}  \C_{\mathbf{i},\downarrow,1}\right)\\
 &
 =  -\delta_{\mathbf{i}\mathbf{j}}\Cd_{\mathbf{j},\uparrow,1}\Cd_{\mathbf{j},\downarrow,2}\C_{\mathbf{i},\uparrow,2}  \C_{\mathbf{i},\downarrow,1},
\end{align*}
where we have made use of $\left[ A,BC\right] = \{A,B\}C-B\{A,C\}$.
Furthermore, $\left[\B_\mathbf{i},\C_\mathbf{i,\sigma,1}\right]=0$, $\B_\mathbf{i}\C_{\mathbf{i},\sigma,1}=0$ and that $\n_{\mathbf{i},1}\C_{\mathbf{i},1}=0$.
With these results, we can calculate
\begin{equation}
    \begin{aligned}
    \hat{\Tilde{c}}_{\mathbf{i}\sigma} \hat{\Tilde{c}}_{\mathbf{j}\sigma^\prime}^\dagger &= \Bd_\mathbf{i} \C_{\mathbf{i},\sigma,1}\Cd_{\mathbf{j}\sigma^\prime,1} \B_\mathbf{j}\notag\\ & = \Bd_\mathbf{i}\left[\delta_{\mathbf{i}\mathbf{j}}(\delta_{\sigma \sigma^\prime}(1-\n_{\mathbf{i},1})-\Cd_{\mathbf{i}\sigma^\prime,1}\C_{\mathbf{i},\sigma,1})-\Cd_{\mathbf{j}\sigma^\prime,1}\C_{\mathbf{i},\sigma,1}\right]\B_\mathbf{j} \notag\\ &= \delta_{\mathbf{i}\mathbf{j}}\delta_{\sigma \sigma^\prime}\Bd_\mathbf{i}\B_\mathbf{i} - \Cd_{\mathbf{j},\sigma^\prime,1}\Bd_\mathbf{i}\B_\mathbf{j}\C_{\mathbf{i},\sigma,1} \notag \\
    &= \delta_{\mathbf{i}\mathbf{j}}\left[\delta_{\sigma \sigma^\prime}\Bd_\mathbf{i}\B_\mathbf{i} + \Cd_{\mathbf{i},\sigma^\prime,1}\C_{\mathbf{i},\sigma,1}(1-\n_{\mathbf{i},2})\right] - \hat{\Tilde{c}}_{\mathbf{j}\sigma^\prime}^\dagger\hat{\Tilde{c}}_{\mathbf{i}\sigma} ,
    \label{eq:newc_anticommutation_v1}
    \end{aligned}
\end{equation}
where we have used that $\n_{\mathbf{i},1}\B_{\mathbf{i}}=0$, $\C_{\mathbf{i},\sigma,1}\B_{\mathbf{i}}=0$ (2nd to 3rd line), and  $\n_{\mathbf{i},1}\C_{\mathbf{i},\sigma,1}=0$ and $\B_{\mathbf{i}}\C_{\mathbf{i},\sigma,1}=0$ (3rd to 4th line). In particular, Eq. \eqref{eq:newc_anticommutation} implies
$$
\langle \holesigma \vert \hat{\Tilde{c}}_{\mathbf{i}\sigma} \hat{\Tilde{c}}_{\mathbf{i}\sigma}^\dagger \vert \holesigma\rangle = 0+ 1-1=0
$$
and 
$$
\langle \singlet \vert \hat{\Tilde{c}}_{\mathbf{i}\uparrow} \hat{\Tilde{c}}_{\mathbf{i}\uparrow}^\dagger \vert \singlet\rangle = 1+(\frac{1}{2}-\frac{1}{2}) =1,
$$
as we expect for fermions. From Eq. \eqref{eq:newc_anticommutation_v1} we obtain Eq. \eqref{eq:newc_anticommutation}.

\subsection{Derivation of $\Ham_\mathrm{eff}$}
In order to derive Eq. \eqref{eq:Bilayer_to_tJ}, we start from the mixed-dimensional bilayer $t$-$J$ model, and explicitly consider different in-plane hoppings and superexchange interactions, $t_\parallel^\alpha$ and $J_\parallel^\alpha$ respectively,
\begin{equation}
\begin{aligned}
    \Ham = \mathcal{P} \left[-\sum_\alpha t_\parallel^\alpha \sum_{\langle \mathbf{i},\mathbf{j}\rangle}\sum_\sigma \left( \Cd_{\mathbf{i},\sigma,\alpha} \C_{\mathbf{j},\sigma,\alpha} + \hc \right) \right] \mathcal{P}
     +\sum_\alpha J_\parallel^\alpha &\sum_{\langle \mathbf{i},\mathbf{j}\rangle} \left( \mathbf{S}_{\mathbf{i},\alpha} \mathbf{S}_{\mathbf{j},\alpha} - \frac{1}{4}\n_{\mathbf{i},\alpha}\n_{\mathbf{j},\alpha} \right)  \\ &+J_\perp \sum_\mathbf{i}  \left(\mathbf{S}_{\mathbf{i},1} \mathbf{S}_{\mathbf{i},2}- \frac{1}{4}\n_{\mathbf{i},1}\n_{\mathbf{i},2} \right),
    \label{eq:mixDtJ}
\end{aligned}
\end{equation}
with $J_\parallel^\alpha=4t^2/U_\alpha$ and $\alpha=1,2$. The hopping term $t_\parallel^{\alpha=1}$ vanishes since we consider a fully occupied upper layer $\alpha=1$. The hopping term in layer $\alpha=2$ leads to indirect hopping of singlets (or singly occupied rungs respectively), 
\begin{align*}
     \bra{\singlet\holeup} \Ham_{t_\parallel^{\alpha=2}} \ket{\holeup\singlet} =  \frac{t_\parallel^{\alpha=2}}{\sqrt{2}} \bra{\singlet\holeup} \left(\ket{\upup \holedown} - \ket{\downup \holeup}\right)   = -\frac{t_\parallel^{\alpha=2}}{\sqrt{2}}\braket{\singlet\holeup | \downup \holeup} = \frac{t_\parallel^{\alpha=2}}{2},
\end{align*}
and gives rise to the first term in Eq. \eqref{eq:Bilayer_to_tJ}. The $J_\parallel^{\alpha=2}$ spin interaction term vanishes since there are either empty sites or singlets, and only 
$$-\frac{J_\parallel^{\alpha=2}}{4}\n_{\mathbf{i},2}\n_{\mathbf{j},2}= -\frac{J_\parallel^{\alpha=2}}{4}\Bd_{\mathbf{i}}\B_{\mathbf{i}}\Bd_{\mathbf{j}}\B_{\mathbf{j}} = -\frac{J_\parallel^{\alpha=2}}{4}(1-\hat{\Tilde{n}}_{\mathbf{i}})(1-\hat{\Tilde{n}}_{\mathbf{j}})$$
remains, since in the low energy subspace the spin  interaction term gives only contributions if no singlets are involved. The same holds for the upper layer $\alpha=1$, where this is taken into account by the new spin operators
\begin{align}
    \hat{\Tilde{\mathbf{S}}}_{\mathbf{i}} &= \hat{\Tilde{c}}^\dagger_{\mathbf{i},\mu}\frac{\sigma_{\mu \nu}}{2} \hat{\Tilde{c}}_{\mathbf{i}\nu} = \Cd_{\mathbf{i},\mu, 1}\B_\mathbf{i} \frac{\sigma_{\mu \nu}}{2} \Bd_\mathbf{i} \C_{\mathbf{i},\nu,1} = \Cd_{\mathbf{i},\mu,1} \frac{\sigma_{\mu\nu}}{2} (1-\n_{\mathbf{i},2})\C_{\mathbf{i},\nu,1} \notag \\
    &=  \Cd_{\mathbf{i},\mu,1} \frac{\sigma_{\mu\nu}}{2} \C_{\mathbf{i},\nu, 1}(1-n_{\mathbf{i},2}) =\hat{{\mathbf{S}}}_{\mathbf{i},1}(1-\Bd_\mathbf{i}\B_\mathbf{i}),
\end{align}
where again we have made use of $\B_\mathbf{i}\C_{\mathbf{i},\sigma,1}=0$ and $\n_{\mathbf{i},1}\C_{\mathbf{i},\nu,1}=0$.
Lastly, there is a constant $-J_\parallel^{\alpha=1}/4$ contribution in Eq. \eqref{eq:mixDtJ} since $\n_{\mathbf{i},1}\n_{\mathbf{j},1}=1$ everywhere. This yields
\begin{align}
    \Ham_\mathrm{eff} = \mathcal{P}\left[ -\frac{t_\parallel^{\alpha=2}}{2}\sum_{\langle \mathbf{i},\mathbf{j}\rangle}\hat{\Tilde{c}}^\dagger_{\mathbf{i}\sigma}\hat{\Tilde{c}}_{\mathbf{j}\sigma} + \hc \right]\mathcal{P}
    +  \sum_{\langle \mathbf{i},\mathbf{j} \rangle} \left[ J_\parallel^{\alpha=1} \left(\Tilde{\mathbf{S}}_\mathbf{i} \cdot \Tilde{\mathbf{S}}_\mathbf{j} -\frac{1}{4}\right)-\frac{J_\parallel^{\alpha=2}}{4}\hat{\Tilde{n}}_{i}^h \hat{\Tilde{n}}_{j}^h\right].
\end{align}

In the derivation of the $t-J$ mixD model Eq. \eqref{eq:mixDtJ}, an additional three-site term
\begin{align}
    \Ham_{3s}^\alpha=-\frac{J_\parallel^{\alpha}}{4}\sum_{\langle i,j,k\rangle}\sum_\sigma \left(\Cd_{\mathbf{k},\sigma,\alpha} \n_{\mathbf{j},\Bar{\sigma},\alpha} \C_{\mathbf{i},\sigma,\alpha}
    + \Cd_{\mathbf{k},\sigma,\alpha} \Cd_{\mathbf{j},\Bar{\sigma},\alpha}\C_{\mathbf{j},\sigma,\alpha} \C_{\mathbf{i},\Bar{\sigma},\alpha} +\hc\right)
    \label{eq:3site}
\end{align}
appears that is often neglected. This term gives only a contribution for the doped layer $\alpha=2$. Furthermore, for large $J_\perp$ the second term of Eq. \eqref{eq:3site} flips the spin at site $\mathbf{j}$ and hence projects out of the rung-singlet low-energy subspace -- a process that we neglect to first order in $J_\parallel^\alpha$. The first term leads to a next-nearest neighbor hopping of dopants in the effective model if the intermediate site $\mathbf{j}$ is empty,
\begin{align}
    \Ham_{\mathrm{eff},3s}=-\frac{J_\parallel^{\alpha=2}}{8}\sum_{\langle i,j,k\rangle}\sum_\sigma \left(\Hat{\Tilde{c}}^\dagger_{\mathbf{k},\sigma} \Hat{\Tilde{n}}^h_{\mathbf{j},\Bar{\sigma}} \Hat{\Tilde{c}}_{\mathbf{i},\sigma}
     +\hc\right).
\end{align}

\twocolumngrid
\bibliographystyle{apsrev4-1}
\bibliography{bilayer}

\end{document}